\documentclass[final,1p,times,authoryear]{elsarticle}

\usepackage{amssymb}
\usepackage{amsmath}

\usepackage[english]{babel}
\usepackage{inconsolata} 
\usepackage{booktabs} 
\usepackage{multirow} 
\usepackage{arydshln} 
\usepackage{fvextra}  
\usepackage[table,dvipsnames]{xcolor} 
\usepackage{placeins} 
\usepackage[hyphens]{url}
\usepackage{textpos} 

\usepackage{tikz} 
\usetikzlibrary{arrows.meta, plotmarks, positioning, shapes.misc, shapes.geometric, quotes, calc}

\graphicspath{{./img/}}

\makeatletter
\def\ps@pprintTitle{   \let\@oddhead\@empty
   \let\@evenhead\@empty
   \def\@oddfoot{\reset@font\hfil\thepage\hfil}
   \let\@evenfoot\@oddfoot
}
\makeatother

\journal{Ecological Modelling}

\begin{document}

\begin{frontmatter}

\title{Can Large Language Models Implement Agent-Based Models? An ODD-based Replication Study}

\author[label1,label2]{Nuno Fachada}
\author[label3,label4]{Daniel Fernandes}
\author[label1,label2]{Carlos M. Fernandes}
\author[label5]{João P. Matos-Carvalho}

\affiliation[label1]{organization={Lusófona University},
            addressline={Campo Grande, 376},
            city={Lisboa},
            postcode={1749-024},
            country={Portugal}}

\affiliation[label2]{organization={INESC INOV-Lab},
            city={Lisboa},
            postcode={1000-029},
            country={Portugal}}

\affiliation[label3]{organization={ISCTE--Instituto Universitário de Lisboa},
            city={Lisboa},
            postcode={1649-026},
            country={Portugal}}

\affiliation[label4]{organization={Instituto de Telecomunicações},
            city={Lisboa},
            postcode={1049-001},
            country={Portugal}}

\affiliation[label5]{organization={LASIGE, Departamento de Informática, Faculdade de Ciências, Universidade de Lisboa},
    city={Lisboa},
    postcode={1749-016},
    country={Portugal}
}

\begin{abstract}
Large language models (LLMs) can now synthesize non-trivial executable code from textual descriptions, raising an important question: can LLMs reliably implement agent-based models from standardized specifications in a way that supports replication, verification, and validation? We address this question by evaluating 17 contemporary LLMs on a controlled ODD-to-code translation task, using the PPHPC predator-prey model as a fully specified reference. Generated Python implementations are assessed through staged executability checks, model-independent statistical comparison against a validated NetLogo baseline, and quantitative measures of runtime efficiency and maintainability. Results show that behaviorally faithful implementations are achievable but not guaranteed, and that executability alone is insufficient for scientific use. GPT-4.1 consistently produces statistically valid and efficient implementations, with Claude 3.7 Sonnet performing well but less reliably. Overall, the findings clarify both the promise and current limitations of LLMs as model engineering tools, with implications for reproducible agent-based and ecological modeling.

\end{abstract}

\begin{keyword}
natural language model specification \sep
specification-to-code translation \sep
code generation \sep
computational reproducibility \sep
verification and validation

\end{keyword}

\end{frontmatter}

\begin{textblock*}{190mm}(-3cm,-18.18cm)
    {\noindent \footnotesize \color{black!90} The peer-reviewed version of this paper is
    published in Ecological Modelling at \url{https://doi.org/10.1016/j.ecolmodel.2026.111624}.
    This version is typeset by the authors and differs only in pagination and
    typographical detail.}
\end{textblock*}

\section{Introduction}
\label{sec:intro}

Agent-based modeling (ABM) is widely used to study complex environmental and social-ecological systems in which macro-level patterns emerge from heterogeneous, localized interactions among individual entities and their environment~\citep{filatova2013spatial,sun2016simple,schulze2017agent}. ABMs are frequently employed to represent coupled human-natural processes and spatially explicit decision-making, while also highlighting the methodological challenges that arise from model flexibility, data demands, and computational cost~\citep{filatova2013spatial,sun2016simple}. Since ABMs can encode rich behavioral rules and fine-grained interaction structure, they can support both explanatory analysis and scenario exploration. However, these same features increase sensitivity to implementation choices and raise the bar for transparent communication of model structure and experiments~\citep{schulze2017agent}.

For simulation models, and ABMs in particular, replication and validation are central to scientific credibility and cumulative knowledge building. Replication helps detect implementation errors, clarify underspecified assumptions, and assess the robustness of conclusions to software platforms, programming languages, and stochastic mechanisms~\citep{thiele2015replicating,donkin2017replicating}. At the same time, the verification and validation perspective emphasizes that simulation results are only as trustworthy as the evidence that the implemented model is correct with respect to its specification (verification) and adequate for its intended purpose (validation)~\citep{david2017vvs}. In computational science more broadly, reproducibility is often framed as a practical minimum standard for evaluating published claims when fully independent replication is costly or infeasible, further motivating workflows that expose code, data, and precise experimental conditions~\citep{peng2011reproducible}. For ABMs, these concerns are amplified by stochasticity, asynchronous interactions, and the prevalence of informal textual specifications, which can leave key implementation details ambiguous and hinder cross-implementation comparability~\citep{donkin2017replicating,david2017vvs}.

A major community response to the documentation and replication challenge has been the ODD (Overview, Design concepts, Details) protocol for describing agent-based models~\citep{grimm2006standard}. ODD introduced a standardized structure intended to make ABM descriptions more complete, readable, and comparable, thereby supporting model understanding and reimplementation. Subsequent revisions addressed limitations observed in practice: an initial update refined some elements of the protocol and clarified its use in publications~\citep{grimm2010odd}, while a second update further improved guidance and suitability for complex models, explicitly emphasizing replication and structural realism~\citep{grimm2020odd}. In parallel, several extensions have been proposed; notably, ODD+D augments the protocol to better represent human decision-making, reflecting the growing prominence of social and social-ecological modeling applications~\citep{muller2013describing}. Most recently, work integrating ODD usage with practical replication workflows (including NetLogo-based replication) highlights both the continuing relevance of standardized descriptions and the persistent effort required to translate textual specifications into faithful implementations~\citep{grimm2025using,berger2024towards}.

Recent advances in large language models (LLMs) have enabled systems capable of generating executable programs directly from textual descriptions~\citep{jiang2026survey}. This capability has prompted growing interest in using LLMs to automate software-intensive aspects of scientific computing, including code synthesis, translation, and interaction with computational workflows through natural-language specifications~\citep{dhruv2025leveraging,yildiz2025large}. Empirical studies further show that LLMs can generate correct and executable code for nontrivial scientific and engineering tasks under zero-shot prompting~\citep{fernandes2025deepseek,fachada2025gpt41}. More broadly, LLMs are increasingly explored as general-purpose assistants throughout the research lifecycle, including experiment design, analysis, and computational reasoning~\citep{luo2025llm4srsurvey,charness2025next}. At the same time, there are persistent concerns regarding correctness, reproducibility, and the fidelity with which LLM-generated code reflects intended model specifications rather than merely producing runnable artifacts~\citep{pearce2025asleep}.

In this paper, a heterogeneous set of modern LLMs is evaluated on a controlled specification-to-code translation task: generating self-contained Python implementations of an agent-based model from a complete ODD specification, under a fixed prompting and execution protocol, and validating the resulting implementations against a NetLogo~\citep{wilensky1999netlogo} baseline using a model-independent statistical output comparison methodology. As the test case, this study uses the PPHPC (Predator-Prey for High-Performance Computing) model, a spatial agent-based predator-prey system explicitly designed as a reference model for implementation, replication, and performance studies in agent-based modeling~\citep{fachada2015template}. The PPHPC model is fully specified and statistically characterized, with the explicit goal of enabling reliable comparison over independent implementations, programming languages, and computational strategies, and it is documented using the ODD protocol in its post-first update form~\citep{grimm2010odd,fachada2015template}. The model exhibits nontrivial stochastic dynamics driven by local agent interactions, a discrete-time schedule with asynchronous state updates within iterations, and probabilistic reproduction, while remaining conceptually simple. These properties make the PPHPC model well suited for assessing whether LLM-generated code not only executes, but also reproduces intended population-level dynamics under statistical validation.

The study aims to clarify not only whether generated code executes, but whether it reproduces the stochastic output distributions of the reference model and whether successful implementations are computationally practical and maintainable. Two research questions guide our analysis:

\begin{quote}

\textbf{RQ1}: To what extent can contemporary LLMs consistently generate executable Python implementations of an agent-based model from a complete ODD specification that are statistically indistinguishable from a validated baseline?\\

\textbf{RQ2}: Among LLM-generated implementations that successfully reproduce the reference model's output, how do \emph{runtime efficiency} and \emph{maintainability-relevant code properties} (e.g., complexity, maintainability, and static analysis warnings) vary \emph{between LLMs} and across successful trials of the same LLM?

\end{quote}

By combining staged execution outcomes, robust statistical validation, runtime measurements under multiple parameter sets, and static code quality metrics, this study provides a multi-criterion assessment of what can currently be expected from LLMs when implementing ABMs from standardized textual specifications, and highlights implications for reproducible simulation practice. The adequacy of an ABM implementation, however, depends on the purpose for which the model is used~\citep{edmonds2019different}. For replication-oriented uses such as the one considered here, statistical agreement with a validated baseline is a useful criterion. For explanatory or teaching-oriented uses, additional considerations such as mechanistic transparency, internal faithfulness, and code simplicity may also be important.

This paper is organized as follows. Section~\ref{sec:background} contextualizes this paper within prior work on translating textual and natural-language model descriptions into executable simulation models, with particular attention to recent uses of LLMs and to the comparatively limited evidence in the ABM domain. Section~\ref{sec:methods} outlines the experimental design and methodological framework used to evaluate LLM-generated implementations, including the reference model, evaluation pipeline, and analysis procedures. Section~\ref{sec:results} presents the empirical findings that address the research questions, while Section~\ref{sec:discussion} interprets these results in light of the study's objectives and broader implications. Section~\ref{sec:lims} discusses limitations, and Section~\ref{sec:conclusions} concludes by summarizing the main contributions and outlining directions for future work. Since this study is presented in an ABM research context but also draws on concepts from contemporary evaluation of LLM-generated code, some LLM terminology and software analysis is unavoidable; to support non-specialist readers, selected terms are briefly clarified where they first appear, and a glossary of core concepts is provided in \ref{app:glossary}.

\section{Background}
\label{sec:background}

LLMs are increasingly discussed in relation to ABM, but most published work connects the two in a different way than the present study: rather than using LLMs to \emph{generate} ABM implementations from specifications, many contributions focus on using LLMs \emph{inside} ABMs as cognitively richer agents (e.g., for natural-language interaction, planning, and decision-making) or as agent controllers in simulation environments~\citep{gao2024large}. This line of work treats LLMs primarily as components of the modeled system, whereas our focus is on LLMs as a \emph{model engineering} tool for translating a formal specification into executable simulation code.

More broadly, simulation researchers have begun to situate generative artificial intelligence (AI) as a potential accelerator across the simulation research lifecycle. For instance, \citet{akhavan2024generative} outline opportunities and failure modes for using LLMs from early problem formulation through model building, experimentation, and reporting, while emphasizing that LLMs should complement rather than replace domain reasoning and methodological rigor. Similarly, \citet{andelfinger2025intelligent} examine the integration of LLMs into the modeling and simulation life cycle, explicitly identifying natural language to code generation as a promising but still limited pathway toward automating model creation and simulation workflows. In the ABM context, \citet{berger2024towards} note that LLMs can already generate modular code snippets that resemble reusable building blocks, while cautioning that such outputs currently lack the transparency, testing, and verification required for reliable scientific use.

Attempts to translate natural language into executable simulations long predate modern LLMs. Early work explored prototype systems that map textual requirements into simulation models in specific technical formalisms. An early example by \citet{cyre1995generating} describes the automatic generation of VHDL\footnote{Although best known as a hardware description language, VHDL was also historically used as a general-purpose discrete event simulation language; in this context, it served as a simulation formalism rather than as a hardware design target.} simulation models from natural language requirements, illustrating the long-standing appeal of reducing the manual effort required to produce executable models from textual descriptions. Later work continued to explore controlled natural language and domain-specific language constructs to narrow the gap between informal descriptions and simulatable specifications, for example by introducing natural language modeling constructs targeted at production and logistics systems~\citep{mayer2015natural}. These pre-LLM approaches typically relied on constrained grammars, domain ontologies, and explicit intermediate representations, trading generality for interpretability and stronger guarantees.

With the advent of modern LLMs, interest has shifted toward using these models as general-purpose translators from free form descriptions to executable model artifacts. In computational biology, \citet{maeda2023automatic} demonstrate the generation of SBML (Systems Biology Markup Language) kinetic models from complex natural language descriptions, showing that LLM-centered pipelines (though not LLMs in isolation) can produce formal, simulatable representations in established scientific standards. In operations research and discrete event contexts, \citet{jackson2024natural} report that a system built on GPT-3 Codex can generate functionally valid simulations for queuing and inventory management problems from verbal explanations, suggesting that LLM-assisted workflows can synthesize runnable models even when the specification is largely prose. Related work has explored adjacent automation tasks that directly enable simulation execution, such as LLM-driven pipelines for generating molecular dynamics input files \citep{chandrasekhar2025automating}. Other contributions have moved closer to full model synthesis, for example by translating natural language conversations into executable discrete event simulation models through a constrained, template-based modeling framework \citep{elbasheer2025natural}. In parallel, benchmarks have begun to formalize evaluation tasks and datasets for ``natural language to simulation code'' translation; \citet{ahmed2025simcode} introduce SIMCODE to assess LLMs' ability to generate ns-3 network simulation code from natural language (where ns-3 is a widely used discrete event network simulator), reflecting a shift from anecdotal demonstrations toward systematic measurement.

However, within ABM specifically, ``natural language to code'' generation remains comparatively underexplored relative to other simulation paradigms and relative to the dominant ``LLMs as agents'' framing. \citet{frydenlund2024modeler} provide one of the clearest early demonstrations in this direction, examining whether ChatGPT can generate functional simulation model code from a prose narrative describing a real-world change process. Their study highlights both the promise of rapid prototyping and the central challenge that motivates the present work: even when LLMs produce runnable code, it can be difficult to establish whether the implementation faithfully matches the intended model dynamics without explicit validation procedures.

In summary, prior work establishes (i) a long-running interest in translating textual descriptions into executable simulations, (ii) renewed optimism due to LLM code synthesis and emerging evaluation benchmarks in non-ABM domains, and (iii) comparatively limited evidence for ABM-specific specification-to-code translation under rigorous behavioral validation. The present study contributes to this gap by evaluating multiple contemporary LLMs on a controlled ODD-to-Python ABM replication task, assessing success not only by executability but by statistical equivalence to a validated baseline and by practical properties of the generated code.

\section{Methodology}
\label{sec:methods}

This section describes the methodological framework used to evaluate whether current LLMs can translate a formal agent-based model specification into statistically valid and computationally practical implementations. Section~\ref{sec:methods:pphpc} introduces the reference PPHPC model, followed by the evaluated LLMs and their inference configuration in Section~\ref{sec:methods:llms}. Section~\ref{sec:methods:pipeline} describes the execution and validation pipeline applied to generated implementations, while Section~\ref{sec:methods:dataanalysis} defines the evaluation metrics and analysis procedures. Finally, Section~\ref{sec:methods:computenv} summarizes the computational environment in which experiments and analyses were performed.

\subsection{The PPHPC model}
\label{sec:methods:pphpc}

\begin{figure} 
    \centering
    \includegraphics[width=1\textwidth]{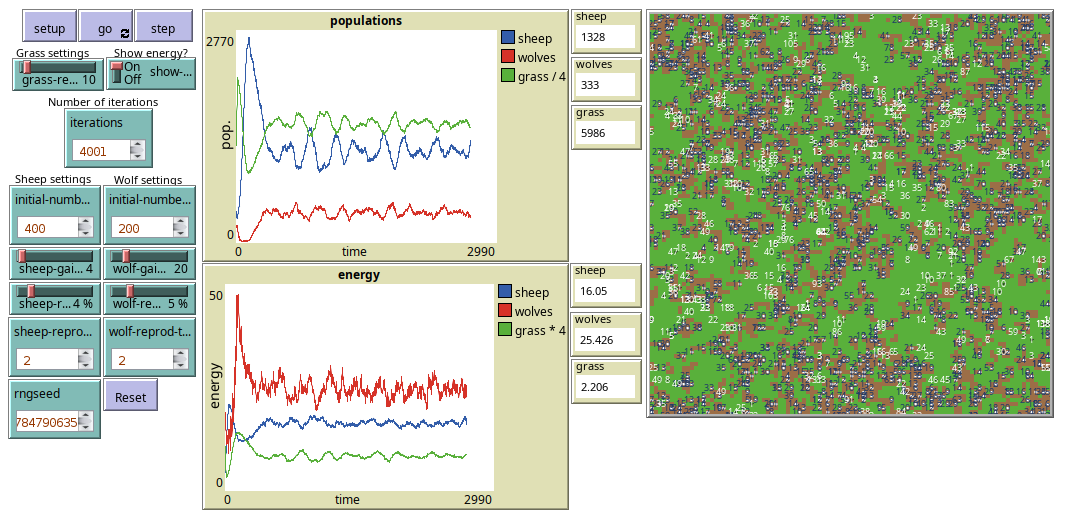}
    \caption{The NetLogo baseline implementation of PPHPC. The interface displays controls for initialization, parameterization, and execution (left), time series of population sizes (prey, predators, and available cell-bound food) and mean agent energies over simulation iterations (center-left), alongside a spatial view of the toroidal grid showing agent and resource distributions (right).}

    \label{fig:pphpc}
\end{figure}

PPHPC is a discrete-time, spatial agent-based predator-prey model defined on a toroidal two-dimensional grid and fully specified using the ODD protocol~\citep{fachada2015template}. The model is conceptually derived from NetLogo's Wolf Sheep Predation model~\citep{wilensky1997wolfsheep}, but was reformulated and formalized with the explicit goals of reproducibility, statistical output analysis, and cross-implementation comparability. In contrast to its pedagogical predecessor, PPHPC is designed as a platform-independent reference model whose dynamics can be faithfully reproduced across heterogeneous implementations. The NetLogo implementation, depicted in Fig.~\ref{fig:pphpc}, is the model's baseline, although validated parallel Java~\citep{fachada2015parallelization} and C/OpenCL~\citep{fachada2017assessing} also exist.

The model comprises three entity types: agents (prey and predators), grid cells containing renewable food resources, and the environment. Agents occupy grid cells, move stochastically within a von Neumann neighborhood, lose energy when moving, gain energy when feeding, and reproduce probabilistically once an energy threshold is exceeded; agents die when their energy reaches zero. Prey consume cell-bound food that regenerates after a fixed countdown, whereas predators consume prey.

Each simulation iteration follows a fixed global schedule consisting of agent movement, food regeneration, agent actions, and output collection. During the action phase, agents are processed in random order, but each agent executes its internal actions in a fixed sequence---first attempting to feed, and then attempting to reproduce---ensuring deterministic intra-agent behavior while preserving stochastic interaction effects. State variables are updated immediately as actions occur, yielding asynchronous state updates within an otherwise synchronous iteration structure. Following the reference specification, simulations are run for 4000 iterations and evaluated under two predefined dynamics-related parameter sets~\citep{fachada2015template}. For a grid of $100\times100$ cells, simulations start with 400 prey and 200 predators; relative to parameter set~1, parameter set~2 typically yields denser populations and higher computational load due to the following changes: an increase in prey energy gain (4 $\rightarrow$ 30), prey reproduction probability (4 $\rightarrow$ 10), and cell food restart (10 $\rightarrow$ 15), together with a decrease in predator energy gain (20 $\rightarrow$ 10); all other parameters remain unchanged. Model output consists of time series of population sizes (prey, predators, and available food) and mean agent energies, which are used as the basis for statistical comparison between implementations. Fig.~\ref{fig:pphpc} exemplifies the dynamics of these time series for parameter set~1.

\subsection{Evaluated LLMs and Inference Configuration}
\label{sec:methods:llms}

The set of language models evaluated in this work (Table~\ref{table:llms}) was chosen to span modern LLM architectures that are potentially relevant when translating an ABM specification into an executable implementation, including (i) general-purpose versus code-specialized training, (ii) dense versus mixture-of-experts (MoE) architectures, (iii) short- versus long-context inference, and (iv) proprietary hosted access versus locally deployable open weight execution. In addition to architectural diversity, the selection intentionally spans distinct provider ecosystems and deployment regimes, reducing the risk that observed outcomes are driven by idiosyncratic training data, engineering choices, or platform-specific behaviors of any single model family. This breadth is particularly relevant for PPHPC, whose dynamics are sensitive to implementation details and therefore stress both faithful interpretation of a textual specification and accurate code generation.

\newcommand{\llmonline}{Hosted} 
\newcommand{\llmoffline}{Local} 

\begin{table}[!tb]
    \centering
    \footnotesize

\caption{LLMs evaluated in this study. Columns indicate the model family, model identifier, release patch or release date (``Patch/Date''), reference publication (``Ref.''), parameter count (``Params.''; a rough indicator of model size), maximum context window in tokens (``Context''; the maximum amount of text the model can consider at once), and execution mode (``Mode''). Parameter counts and context sizes are shown with magnitudes indicated directly in the table (K $\rightarrow$ thousands, M $\rightarrow$ millions, B $\rightarrow$ billions). ``Mode'' indicates whether the model was run locally via Ollama on institutional infrastructure or accessed remotely (hosted) through the provider.}

    \label{table:llms}

\begin{tabular}{llllrrl}
    \toprule
    Family & Model & Patch/Date & Ref. & Params. & Context & Mode \\

    \midrule

    Claude &
    \texttt{claude-3.7-sonnet} &
    20250219 &
    \cite{anthropic2025claude37} &
    $\star$ &
    200 K &
    \llmonline  \\

    \addlinespace

    DeepSeek &
    \texttt{deepseek-coder-v2} &
    2024-09-06 &
    \cite{zhu2024deepseek} &
    16 B &
    128 K &
    \llmoffline \\

    &
    \texttt{deepseek-r1} &
    2025-01-20 &
    \cite{guo2025deepseekr1} &
    671 B &
    128 K &
    \llmonline \\

    &
    \texttt{deepseek-v3} &
    0324 &
    \cite{liu2024deepseekv3} &
    671 B &
    128 K &
    \llmonline \\

    \addlinespace

    Gemma &
    \texttt{codegemma} &
    2024-07-18 &
    \cite{zhao2024codegemma} &
    7 B &
    8 K &
    \llmoffline  \\

    &
    \texttt{gemma3} &
    2025-04-18 &
    \cite{kamath2025gemma3} &
    27 B &
    128 K &
    \llmoffline \\

    \addlinespace

    GPT &
    \texttt{gpt-4o} &
    2024-08-06 &
    \cite{hurst2024gpt} &
    $\star$ &
    128 K &
    \llmonline  \\

    &
    \texttt{gpt-4.1} &
    2025-04-14 &
    \cite{openai2025gpt41} &
    $\star$ &
    1 M &
    \llmonline \\

    \addlinespace

    Grok &
    \texttt{grok-3-beta} &
    2025-02-17 &
    \cite{xai2025grok3beta} &
    $\star$ &
    1 M &
    \llmonline \\

    \addlinespace

    LLaMA &
    \texttt{llama3.3} &
    2024-12-06 &
    \cite{grattafiori2024llama} &
    70 B &
    128 K &
    \llmoffline   \\

    &
    \texttt{codellama} &
    2024-07-18 &
    \cite{roziere2023code} &
    70 B &
    100 K &
    \llmoffline \\

    \addlinespace

    Mistral &
    \texttt{codestral} &
    2024-09-03 &
    \cite{mistral2024codestral} &
    22 B &
    32 K &
    \llmoffline \\

    &
    \texttt{mistral-large} &
    24.11 &
    \cite{mistral2024large} &
    123 B &
    128 K &
    \llmonline  \\

    \addlinespace

    Olmo &
    \texttt{olmo2} &
    2025-01-11 &
    \cite{olmo20242} &
    13 B &
    4 K &
    \llmoffline \\

    \addlinespace

    Phi &
    \texttt{phi4} &
    2025-01-08 &
    \cite{abdin2024phi} &
    14 B &
    16 K &
    \llmoffline \\

    \addlinespace

    Qwen &
    \texttt{qwen2.5-coder} &
    2025-05-28 &
    \cite{hui2025qwen25coder} &
    32 B &
    131 K &
    \llmoffline \\

    &
    \texttt{qwen3} &
    2025-05-29 &
    \cite{yang2025qwen3} &
    32 B &
    33 K &
    \llmoffline \\

        \bottomrule
        \multicolumn{5}{l}{\footnotesize{\textsuperscript{$\star$}Not publicly disclosed.}}

    \end{tabular}
\end{table}

The proprietary, hosted models consist of Claude~3.7~Sonnet~\citep{anthropic2025claude37}, GPT-4o~\citep{hurst2024gpt}, GPT-4.1~\citep{openai2025gpt41}, Grok 3~\citep{xai2025grok3beta}, DeepSeek-R1~\citep{guo2025deepseekr1}, DeepSeek-V3~\citep{liu2024deepseekv3}, and Mistral Large~2.1~\citep{mistral2024large}. These models represent recent state-of-the-art systems with strong instruction following and code synthesis capabilities. Some, like for example GPT-4.1 and Grok 3, offer long contexts intended to accommodate lengthy specifications and iterative refinement within a single session. The DeepSeek models additionally provide a contrast between a large MoE general model (V3) and a reasoning-oriented variant (R1), enabling assessment of whether explicitly optimized reasoning translates into more reliable implementations under a fixed prompting protocol.

The locally executed open weight models include DeepSeek Coder-V2~\citep{zhu2024deepseek}, CodeGemma~\citep{zhao2024codegemma}, Gemma~3~\citep{kamath2025gemma3}, Llama~3.3~\citep{grattafiori2024llama}, CodeLlama~\citep{roziere2023code}, Codestral~\citep{mistral2024codestral}, Olmo~2~\citep{olmo20242}, Phi-4~\citep{abdin2024phi}, Qwen2.5-Coder~\citep{hui2025qwen25coder}, and Qwen3~\citep{yang2025qwen3}. This subset emphasizes reproducibility and methodological transparency (especially for Olmo~2), while still covering both generalist and code-oriented variants (e.g., Llama~3.3 vs. CodeLlama; Gemma~3 vs.\ CodeGemma; Qwen3 vs.\ Qwen2.5-Coder). Together with the hosted models, this selection provides a compact but representative set of modern LLM capabilities and deployment regimes for specification-to-code translation in Python.

All LLMs were evaluated under a uniform inference protocol designed to balance limited stochastic exploration with overall coding reliability. Each model was queried six times using the prompt presented in Table~\ref{tab:prompt}, with each query issued in a fresh session so that no conversational context from previous attempts could influence the response. Since LLMs are stochastic generators---especially when sampling temperature (which controls how variable or random the generated text is) is greater than zero---querying each model multiple times allows us to probe that stochasticity and assess not only whether a model can produce a successful implementation, but also how consistently it does so. The prompt was designed to instantiate a controlled specification-to-code translation task by embedding the full ODD model description and constraining the required interface, libraries, and output structure. This reduces underspecification and prompt-induced variability, helping ensure that differences in outcomes primarily reflect model capabilities.

\begin{table}[!htb]
    
    \caption{Standardized prompt used in all LLM trials. The ODD description of PPHPC in \LaTeX markup is omitted for brevity; the full prompt is available in the supplementary material~\citep{fachada2026suppl}.}

    \label{tab:prompt}
    \centering
    \begin{tabular}{l}

        \toprule

\begin{minipage}{0.97\linewidth}
\begin{Verbatim}[fontsize=\scriptsize,breaklines,breaksymbolleft={}, breaksymbolright={}, breakindent=0pt]
Consider the PPHPC agent-based model, described using the ODD protocol as follows:

```latex
{{ODD HERE}}
```

Create a Python function named `run_pphpc()` which implements and simulates the PPHPC model. The function should receive the simulation parameters (all integers) in the following order:

1. Horizontal grid size.
2. Vertical grid size.
3. Initial number of prey.
4. Initial number of predators.
5. Number of iterations to perform.
6. Prey energy gain from food.
7. Predator energy gain from food.
8. Prey energy loss per turn.
9. Predator energy loss per turn.
10. Prey reproduction threshold.
11. Predator reproduction threshold.
12. Prey reproduction probability.
13. Predator reproduction probability.
14. Cell food restart.

The function should return a Pandas dataframe with the simulation output, where each row corresponds to the output at the current iteration, and the columns are as follows:

1. `total_prey`: total prey population count.
2. `total_predators`: total predator population count.
3. `total_food`: quantity of available cell-bound food.
4. `mean_energy_prey`: mean energy of prey population.
5. `mean_energy_predators`: mean energy of predator population.
6. `mean_c`: mean value of the $C$ state variable in all grid cells.
  
The `run_pphpc()` must additionally follow these strict requirements:

- The function should be Python 3.9+ compatible and follow the PEP 8 style guidelines.
- The function must be self-contained, including all necessary imports, constants, and helper code within the function body.
- The function can only make use of the numpy and pandas libraries, as well as any module from the Python standard library.
- The function should not include print statements, plots, or additional output.
\end{Verbatim}
\end{minipage} \\

    \bottomrule
    \end{tabular}
\end{table}

Seed control was used on a best effort basis, acknowledging that strict reproducibility cannot be guaranteed due to model-side and infrastructure-side sources of non-determinism (e.g., parallel execution, hardware heterogeneity, and opaque provider-side updates). Consequently, the study emphasizes statistical consistency throughout generation trials rather than exact reproducibility of individual LLM responses. Sampling temperature was set to a low value ($T=0.1$ or the closest supported equivalent) to reduce variability while still allowing some diversity in generated code. Other sampling parameters (e.g., \texttt{top\_p}) were left at provider defaults.

Offline models were executed locally on the authors' infrastructure (see Section~\ref{sec:methods:computenv}), while proprietary models were accessed via their official APIs. For models that support external tools (e.g., web search or code execution), these capabilities were disabled, ensuring that all implementations were derived solely from the provided ODD specification. This configuration isolates each model's intrinsic ability to translate a natural language ABM description into executable code.

\subsection{Execution and Validation Pipeline}
\label{sec:methods:pipeline}

Fig.~\ref{fig:pipeline} summarizes the execution, evaluation, and scoring pipeline applied to each LLM-generated response. For each model and generation trial, a fixed prompt containing the complete \LaTeX{} ODD specification of PPHPC is combined with a trial-specific random seed and submitted to the LLM. The resulting response is first inspected to determine whether it contains executable Python code exposing the required entry-point function; responses that fail this check are assigned score~1.

\definecolor{incol}{HTML}{fdedd7} 
\definecolor{decol}{HTML}{d6f9fd}
\definecolor{outcol}{HTML}{fef8c8}
\definecolor{opcol}{HTML}{fae3e2}

\definecolor{tincol}{HTML}{dd6126} 
\definecolor{tdecol}{HTML}{479ab7}
\definecolor{toutcol}{HTML}{c38b2b}
\definecolor{topcol}{HTML}{e5958f}

\begin{figure}[!tbp]
    \centering

\resizebox{\textwidth}{!}{%
\begin{tikzpicture}[
    text=black!80,
    conn/.style = {
        ->,
        >={Stealth[round]},
        thick,
        rounded corners,
        black!50,
        text=black!90
    },
    conn_dec/.style = {
        pos=0,
        font=\small,    
    },
    conn_dec_h/.style = {
        conn_dec,
        above,
        xshift=2.5mm
    },
    conn_dec_v/.style = {
        conn_dec,
        right,
        yshift=-2mm
    },
    conn_score/.style = {
        black!50,
        below,
        font=\small
    },
    input/.style = {
        rounded rectangle,
        minimum size=8mm,
        align=center,
        text width=19mm,
        very thick,
        draw=tincol,
        top color=white,
        bottom color=incol,
        font=\small,
        yshift=1cm,
        text height=1.5ex,
        text depth=.25ex
        },
    output/.style = {
        rounded rectangle,
        align=center,
        minimum size=10mm,
        text width=22mm,
        very thick,
        draw=toutcol,
        top color=white,
        bottom color=outcol,
        font=\ttfamily,
        yshift=-1cm,
        text height=1.5ex,
        text depth=.25ex},        
    operation/.style = {
        rectangle,
        align=center,
        minimum size=10mm,
        text width=22mm,
        very thick,
        text=black!80,
        draw=topcol,
        top color=white,
        bottom color=opcol,
        },   
    decision/.style = {
        diamond,
        align=center,
        text width=12mm,
        minimum size=22mm,
        very thick,
        text=black!80,
        draw=tdecol,
        top color=white,
        bottom color=decol,
        font=\small,
        },           
    ]

\node [operation] (llm) {LLM\\inference};

\node [decision,right=of llm] (hascode) {Has\\code?};
\node [operation,right=of hascode] (qrun) {$1\times$ quick run\\(iters=5)};
\node [decision,right=of qrun] (error) {Error?};
\node [operation,right=of error] (frun) {$30\times$ full runs\\(iters=4000)};
\node [decision,right=of frun] (stat) {$\sim$\\Baseline?};

\node [output,below=of llm] (answer) {answer.txt};
\node [output,below=of frun] (scores) {results.csv};

\node [input,above=of llm, xshift=-1.2cm] (prompt) {Prompt};
\node [input,above=of llm, xshift=1.2cm] (seed) {Seed}; 
\node [input,above=of frun, xshift=-1.2cm] (ps1) {Param. set 1}; 
\node [input,above=of frun, xshift=1.2cm] (ps2) {Param. set 2}; 

\draw[conn]
  (prompt.south)
  |- ($(prompt.south)!0.5!(llm.130)$)
  -| (llm.130);

\draw[conn]
    (seed.south)
    |- ($(seed.south)!0.5!(llm.50)$)
    -| (llm.50);

\draw[conn] (llm) -- (answer);

\draw[conn]
    (answer.east)
    -| ($(answer.east)!0.5!(hascode.west)$)
    |- (hascode.west);

\draw[conn] (hascode) -- (qrun) node[conn_dec_h] {Yes};

\draw[conn] (hascode.south)
    |- node[conn_dec_v] {No} (scores.west)
    node[conn_score,pos=0.95] {score=1};

\draw[conn] (qrun) -- (error);

\draw[conn] (error) -- (frun) node[conn_dec_h] {No};

\draw[conn] (ps1) -| (qrun);

\draw[conn]
    (ps1)
    |- ($(ps1.south)!0.5!(frun.130)$)
    -| (frun.130);

\draw[conn]
    (ps2)
    |- ($(ps2.south)!0.5!(frun.50)$)
    -| (frun.50);

\draw[conn] (frun) -- (stat);

\draw[conn]
    (error)
    |- node[conn_dec_v] {Yes} ($(error.south)!0.5!(scores.130)$)
    -| (scores.130) node[conn_score,pos=0.0] {score=2, 3, or 4};

\draw[conn]
    (stat)
    |- node[conn_dec_v] {No} ($(stat.south)!0.5!(scores.50)$)
    -| (scores.50) node[conn_score,pos=0.0] {score=5};

\draw[conn]
    (stat.east)
    -- ++(1cm,0) node[conn_dec_h] {Yes}
    |-  (scores.east) node[conn_score,pos=0.9] {score=6};;

\end{tikzpicture}
} 

\caption{Execution and validation pipeline for LLM-generated PPHPC implementations from an ODD-including prompt and trial seed. The LLM response is first checked for the required code/function (score~1 otherwise), then subjected to a short smoke test (5 iterations; param. set~1) to detect syntax, runtime/timeout, or output format failures (scores~2--4, respectively). Surviving implementations are executed in 30 stochastic replications of 4000 iterations under each of two parameter sets, and their outputs are statistically compared against the NetLogo baseline; disagreement for either parameter set yields score~5, while statistical indistinguishability for both yields score~6.}

    \label{fig:pipeline}
\end{figure}
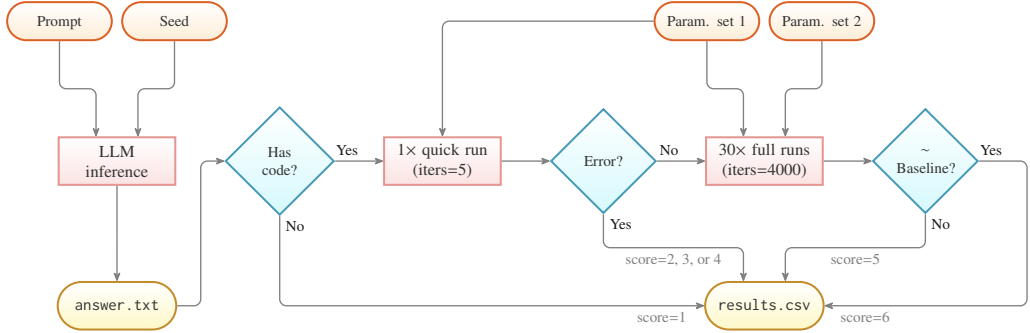

If code is present, the implementation is subjected to a short ``quick run'' consisting of five iterations under parameter set~1. This inexpensive smoke test is used to detect syntax errors (score~2), runtime errors or timeouts (score~3), and violations of the expected output interface or format (score~4) before incurring the cost of full simulation runs. Implementations that pass this stage are then executed in full: 30 independent stochastic replications of 4000 iterations are performed for each of two predefined parameter sets, yielding 60 output trajectories per implementation.

Rather than treating success as a binary outcome, as in pass@$k$-style evaluations~\citep{chen2021evaluating}, the pipeline explicitly records how far each generated implementation progresses through successive stages of execution and validation. In other words, the code is not scored simply as pass or fail; it receives a higher score as it successfully advances through more demanding stages of the pipeline. This staged scoring is motivated by our interest in distinguishing partial successes (e.g., runnable but statistically incorrect models) from complete failures, and by the observation that pass@$k$ provides limited insight into robustness, efficiency, or downstream usability of generated code~\citep{yeo2024framework}.

For implementations that complete all runs, output is compared against a NetLogo reference implementation for distributional equivalence~\citep{axtell1996aligning} based on the model-independent statistical comparison methodology proposed by~\cite{fachada2017model}. Simulation output from each stochastic run is first transformed into a fixed-length feature vector by standard scaling each of the six output time series and concatenating them (run-wise) into a single vector; stacking these vectors yields a 60-row data matrix (30 runs from the LLM-generated implementation and 30 from the NetLogo baseline). This matrix is then projected onto a reduced set of principal components (PCs) capturing at least 80\% of the variance. Distributional equivalence is assessed using the non-parametric, multivariate Energy test~\citep{szekely2013energy} on those PCs, with multiple testing correction performed with the Benjamini-Hochberg procedure~\citep{benjamini1995controlling}. If outputs are statistically indistinguishable from the baseline under both parameter sets, the implementation is assigned score~6; if a significant discrepancy is detected for at least one parameter set, it receives score~5.

For all statistically validated implementations (score~6), execution times for each replication are also recorded. These measurements are used to assess the computational efficiency of correct LLM-generated implementations and to evaluate the practical feasibility of deriving agent-based model code from ODD descriptions using LLMs.

\subsection{Evaluation Metrics and Analysis}
\label{sec:methods:dataanalysis}

This subsection describes the quantitative measures derived from the execution and validation pipeline (Section~\ref{sec:methods:pipeline}) and how they are summarized to answer the research questions.

\subsubsection{Execution outcomes and success rates}

For each LLM, six independent trials are performed (with distinct prompting seeds), yielding an ordinal outcome score in $\{1,\dots,6\}$ per trial. These scores encode how far a generated response progresses through successive stages of code extraction, smoke testing, full execution, and statistical validation. The distribution of scores across the six trials is used to characterize typical failure modes and the stage at which progress most often halts. This directly addresses RQ1 by quantifying the reliability of specification-to-code translation and the extent to which generated implementations reach statistical validity. Since downstream analyses require implementations that are both executable and behaviorally faithful, the primary reliability measure is the \emph{success rate}, defined as the proportion of trials achieving score~6 (statistical indistinguishability from the baseline under both parameter sets). Only successful implementations (score~6) are retained for runtime and static code quality analysis.

\subsubsection{Runtime performance of validated implementations}

For each successful model-trial combination, runtime is recorded for every stochastic replication (30 runs) under each parameter set, and summarized by the mean execution time $\bar{t}$ and its relative standard deviation $s_\mathrm{rel}=100\,s/\bar{t}$, where $s$ is the sample standard deviation across the 30 replications. This evaluation is important because correctness alone does not guarantee predictable computational behavior: even small variations in LLM-generated logic can yield quite different execution costs, ranging from modest slowdowns to severe degradation under load~\citep{rajput2025dynamic}. Although mean runtime is additionally reported relative to the NetLogo baseline (for contextual comparison), this cross-language comparison is not strictly fair since NetLogo and Python execute on different runtimes and incur different constant factors (e.g., interpreter overhead, data structure implementations, and scheduling semantics). 
Consequently, the main performance evidence contributing to RQ2 is derived from \emph{between-implementation} differences among the successful Python implementations, in particular on (i) how runtimes vary between LLMs, and (ii) how much variability is observed over successful trials of the same LLM. To this end, distributions of $\bar{t}$ over successful trials are analyzed per LLM, separately for each parameter set, to assess whether a given LLM tends to generate consistently efficient implementations or exhibits high dispersion in computational cost despite output-level correctness.

\subsubsection{Static code quality of validated implementations}

Assessing static code quality is important since functional correctness alone does not ensure that generated code is maintainable or free of latent structural issues. Here, ``static'' means that the code is examined without executing it. LLM-generated code can vary widely in structural quality, including code smells (recurring patterns that suggest poor design or harder future maintenance), complexity, and implied maintenance effort, even when functionally correct~\citep{sousa2025comparing, sabra2025assessing}. Therefore, for each successful implementation, the following maintainability-relevant static metrics are computed from the generated source code:

\begin{itemize}
    \item \emph{Source lines of code} ($s_\mathrm{loc}$), counted as non-comment lines, to capture verbosity and approximate surface area for comprehension and modification.
    \item \emph{Cyclomatic complexity} ($c_c$), defined as the number of linearly independent control flow paths, as a proxy for branching complexity and the effort required to test and reason about code; lower values are typically preferred~\citep{mccabe1976complexity}.
    \item \emph{Maintainability index} ($m_i$), an empirical composite indicator that summarizes \linebreak maintainability-related properties, including $s_\mathrm{loc}$, $c_c$, percentage of code comments, and operator/operand usage patterns~\citep{oman1992metrics}. We use the version of $m_i$ implemented by~\citet{radon2012}, where the index is normalized to a 0--100 scale and higher values indicate more maintainable code.
    
    \item  \emph{Static type checking error density} ($e_t/100$), defined as the number of type errors normalized per 100~$s_\mathrm{loc}$, to capture potential mismatches between intended and inferred types that may hinder refactoring and increase defect risk.
    \item \emph{Flaws and formatting} issues ($e_F/100$), normalized per 100~$s_\mathrm{loc}$, which include both style- and formatting-related warnings as well as logic, correctness, and dead code problems. This measure captures static warnings commonly associated with likely defects (e.g., unused variables, unused imports, ambiguous variable names).
    
\end{itemize}

These metrics are treated as complementary proxies for maintainability and potential defect risk and, together with runtime performance, address RQ2 by separating behavioral correctness from maintainability-relevant code properties among validated implementations. 

To facilitate comparisons between LLMs while accounting for within-model variability, these metrics are summarized in two complementary ways: (a) per-LLM aggregation using robust location statistics (medians) computed over all successful trials of that LLM; and, (b) visualization of the full distribution of metric values over successful trials, highlighting whether a model tends to produce structurally similar code or diverse implementations even when behaviorally correct.

\subsection{Computational Environment}
\label{sec:methods:computenv}

Three distinct computational environments were used, corresponding to (i) local execution of open weight LLMs, (ii) execution of the NetLogo baseline and the LLM-generated Python implementations for correctness and runtime measurement, and (iii) offline data analysis and figure generation.

Open weight models were prompted and executed via Ollama (version~0.12.3) on \linebreak Ubuntu~22.04.4~LTS, using a workstation equipped with an Intel Xeon Gold~5218 CPU, 128~GB system RAM, and an Nvidia Tesla~T4 GPU (16~GB VRAM). This hardware configuration is reported for completeness, but it is not central to the study, since inference performance for local models is not evaluated and all outcomes are assessed solely from the generated source code.

All simulations used for statistical validation and runtime measurement were executed on a separate machine running Ubuntu~22.04.4~LTS with an AMD Ryzen~7900X CPU and 64~GB system RAM. The baseline implementation was executed using NetLogo~6.4.0-64. LLM-generated implementations were executed in a dedicated Python~3.10.12 virtual environment with \linebreak
NumPy~2.3.2~\citep{harris2020array} and Pandas~2.3.1~\citep{mckinney2011pandas}, reflecting the constrained dependency set specified in the prompt (i.e., generated code could assume only these libraries, in addition to the Python standard library). Since runtime comparisons and scalability relevant behavior depend strongly on hardware and interpreter/library versions, this configuration is treated as part of the experimental context for the performance results presented in Section~\ref{sec:results:runtimeperf}.

Statistical testing, aggregation, and static code quality analysis were performed in a separate Python~3.12.3 virtual environment. Core data wrangling relied on NumPy~2.3.5 and Pandas~2.3.3. Statistical validation used hyppo~0.5.2~\citep{panda2019hyppo} for the multivariate Energy test, with preprocessing based on standard scaling and principal component analysis (PCA) implemented in scikit-learn~1.8.0~\citep{pedregosa2011scikit}. Multiple testing correction was performed with the Benjamini--Hochberg procedure provided by statsmodels~0.14.6~\citep{seabold2010statsmodels}. Static code quality metrics were computed using mypy~1.19.1~\citep{mypy2014} for type checking errors ($e_t$), radon~6.0.1~\citep{radon2012} for source lines of code, cyclomatic complexity, and maintainability index, and ruff~0.14.10~\citep{ruff2022} for flaws and formatting issues ($e_F$). Hardware and operating system details are not considered critical for reproducibility at this analysis stage, since the analysis is deterministic given the recorded outputs and extracted source code; the resulting figures (e.g., the code quality summary plot) were produced under this environment. 

All scripts used to extract and execute generated implementations, run the baseline model, collect simulation outputs and runtime measurements, and perform the statistical and static analyses (including the notebooks used to produce all reported figures and tables), together with the corresponding raw and processed data, are provided in the supplementary material~\citep{fachada2026suppl}.

\section{Results}
\label{sec:results}

\subsection{Execution outcomes and validation}
\label{sec:results:execvalid}

\begin{figure}[!p]
    \centering
    \includegraphics[width=0.76\textwidth]{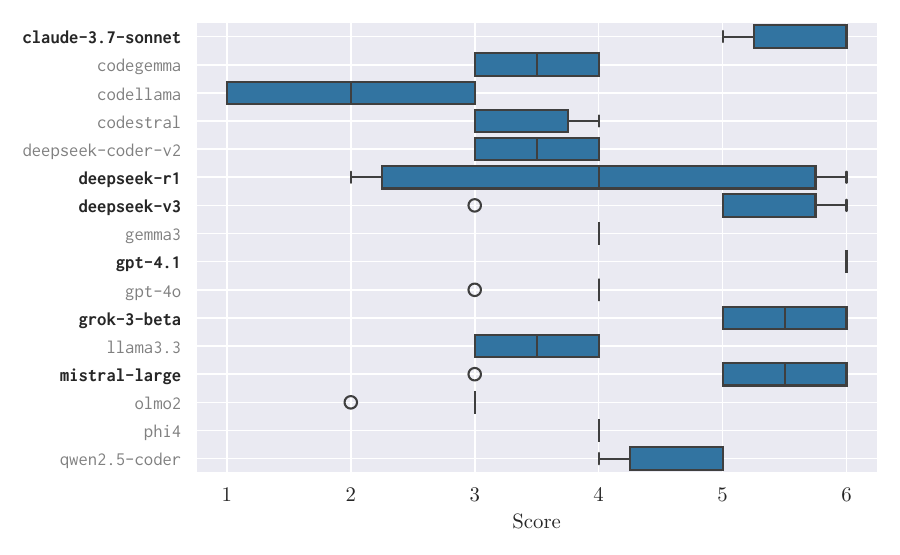}
    \caption{Distribution of implementation outcome scores (1--6) over six trials (different random seeds) for Python code generated by each LLM when implementing the PPHPC simulation model from its ODD protocol description. Higher scores indicate later stages reached in each trial, with a score of 6 denoting success. Models shown in bold achieved success (score = 6) for at least one trial/seed.}

    \label{fig:scores}
\end{figure}

Throughout the six trials, substantial differences were observed in the ability of the evaluated LLMs to generate executable PPHPC implementations that progressed through the full validation pipeline (Table~\ref{tab:results}; Fig.~\ref{fig:scores}). GPT-4.1 consistently produced successful implementations (score~6 in all trials), whereas GPT-4o did not reach statistical validation in any trial, indicating a considerable increase in performance for successive generations of OpenAI models. Among the other hosted models, several achieved success in at least one trial, including Claude~3.7 Sonnet, Mistral Large~2.1, and Grok~3, with DeepSeek-R1 and DeepSeek-V3 displaying less consistent outcomes. 

\definecolor{basecellcol}{gray}{0.6}

\begin{table}[!p]
\centering
\footnotesize
\caption{Execution and validation scores (1–6) for Python code generated by the LLMs listed in the ``Model'' column over six trials (different random seeds) when implementing the PPHPC simulation model from its ODD protocol description. Scores indicate the stage reached in each trial: 1~--~no code/function produced; 2~--~syntax error; 3~--~runtime error or timeout; 4~--~incorrect function output format; 5~--~statistically incorrect simulation results; 6~--~statistically indistinguishable simulation results (success). Column ``Success rate'' reports the percentage of trials achieving score 6. Darker shading indicates higher scores or success rates.}
\label{tab:results}
\begin{tabular}{lcccccccr}
\toprule
 & \multicolumn{6}{l}{Trial/seed} \\
\cmidrule(){2-7}
Model & \multicolumn{1}{r}{1/0088} & \multicolumn{1}{r}{2/0288} & \multicolumn{1}{r}{3/3259} & \multicolumn{1}{r}{4/5613} & \multicolumn{1}{r}{5/8153} & \multicolumn{1}{r}{6/8797} & &  
\multicolumn{1}{l}{Success rate} \\
\midrule

\texttt{claude-3.7-sonnet} & \cellcolor{basecellcol!64} \textcolor{black}{5} & \cellcolor{basecellcol!64} \textcolor{black}{5} & \cellcolor{basecellcol!80} \textcolor{black}{6} & \cellcolor{basecellcol!80} \textcolor{black}{6} & \cellcolor{basecellcol!80} \textcolor{black}{6} & \cellcolor{basecellcol!80} \textcolor{black}{6} & & \cellcolor{basecellcol!56}\textcolor{black}{66.7\%}\\
\texttt{codegemma} & \cellcolor{basecellcol!32} \textcolor{black}{3} & \cellcolor{basecellcol!48} \textcolor{black}{4} & \cellcolor{basecellcol!48} \textcolor{black}{4} & \cellcolor{basecellcol!48} \textcolor{black}{4} & \cellcolor{basecellcol!32} \textcolor{black}{3} & \cellcolor{basecellcol!32} \textcolor{black}{3} & & \cellcolor{basecellcol!0}\textcolor{black}{0.0\%}\\
\texttt{codellama} & \cellcolor{basecellcol!0} \textcolor{black}{1} & \cellcolor{basecellcol!0} \textcolor{black}{1} & \cellcolor{basecellcol!32} \textcolor{black}{3} & \cellcolor{basecellcol!32} \textcolor{black}{3} & \cellcolor{basecellcol!32} \textcolor{black}{3} & \cellcolor{basecellcol!0} \textcolor{black}{1} & & \cellcolor{basecellcol!0}\textcolor{black}{0.0\%}\\
\texttt{codestral} & \cellcolor{basecellcol!48} \textcolor{black}{4} & \cellcolor{basecellcol!32} \textcolor{black}{3} & \cellcolor{basecellcol!32} \textcolor{black}{3} & \cellcolor{basecellcol!32} \textcolor{black}{3} & \cellcolor{basecellcol!48} \textcolor{black}{4} & \cellcolor{basecellcol!32} \textcolor{black}{3} & & \cellcolor{basecellcol!0}\textcolor{black}{0.0\%}\\
\texttt{deepseek-coder-v2} & \cellcolor{basecellcol!48} \textcolor{black}{4} & \cellcolor{basecellcol!32} \textcolor{black}{3} & \cellcolor{basecellcol!48} \textcolor{black}{4} & \cellcolor{basecellcol!32} \textcolor{black}{3} & \cellcolor{basecellcol!32} \textcolor{black}{3} & \cellcolor{basecellcol!48} \textcolor{black}{4} & & \cellcolor{basecellcol!0}\textcolor{black}{0.0\%}\\
\texttt{deepseek-r1} & \cellcolor{basecellcol!80} \textcolor{black}{6} & \cellcolor{basecellcol!16} \textcolor{black}{2} & \cellcolor{basecellcol!32} \textcolor{black}{3} & \cellcolor{basecellcol!16} \textcolor{black}{2} & \cellcolor{basecellcol!80} \textcolor{black}{6} & \cellcolor{basecellcol!64} \textcolor{black}{5} & & \cellcolor{basecellcol!28}\textcolor{black}{33.3\%}\\
\texttt{deepseek-v3} & \cellcolor{basecellcol!64} \textcolor{black}{5} & \cellcolor{basecellcol!64} \textcolor{black}{5} & \cellcolor{basecellcol!64} \textcolor{black}{5} & \cellcolor{basecellcol!32} \textcolor{black}{3} & \cellcolor{basecellcol!80} \textcolor{black}{6} & \cellcolor{basecellcol!80} \textcolor{black}{6} & & \cellcolor{basecellcol!28}\textcolor{black}{33.3\%}\\
\texttt{gemma3} & \cellcolor{basecellcol!48} \textcolor{black}{4} & \cellcolor{basecellcol!48} \textcolor{black}{4} & \cellcolor{basecellcol!48} \textcolor{black}{4} & \cellcolor{basecellcol!48} \textcolor{black}{4} & \cellcolor{basecellcol!48} \textcolor{black}{4} & \cellcolor{basecellcol!48} \textcolor{black}{4} & & \cellcolor{basecellcol!0}\textcolor{black}{0.0\%}\\
\texttt{gpt-4.1} & \cellcolor{basecellcol!80} \textcolor{black}{6} & \cellcolor{basecellcol!80} \textcolor{black}{6} & \cellcolor{basecellcol!80} \textcolor{black}{6} & \cellcolor{basecellcol!80} \textcolor{black}{6} & \cellcolor{basecellcol!80} \textcolor{black}{6} & \cellcolor{basecellcol!80} \textcolor{black}{6} & & \cellcolor{basecellcol!84}\textcolor{black}{100.0\%}\\
\texttt{gpt-4o} & \cellcolor{basecellcol!32} \textcolor{black}{3} & \cellcolor{basecellcol!48} \textcolor{black}{4} & \cellcolor{basecellcol!48} \textcolor{black}{4} & \cellcolor{basecellcol!48} \textcolor{black}{4} & \cellcolor{basecellcol!48} \textcolor{black}{4} & \cellcolor{basecellcol!48} \textcolor{black}{4} & & \cellcolor{basecellcol!0}\textcolor{black}{0.0\%}\\
\texttt{grok-3-beta} & \cellcolor{basecellcol!80} \textcolor{black}{6} & \cellcolor{basecellcol!80} \textcolor{black}{6} & \cellcolor{basecellcol!64} \textcolor{black}{5} & \cellcolor{basecellcol!64} \textcolor{black}{5} & \cellcolor{basecellcol!64} \textcolor{black}{5} & \cellcolor{basecellcol!80} \textcolor{black}{6} & & \cellcolor{basecellcol!42}\textcolor{black}{50.0\%}\\
\texttt{llama3.3} & \cellcolor{basecellcol!32} \textcolor{black}{3} & \cellcolor{basecellcol!48} \textcolor{black}{4} & \cellcolor{basecellcol!32} \textcolor{black}{3} & \cellcolor{basecellcol!48} \textcolor{black}{4} & \cellcolor{basecellcol!48} \textcolor{black}{4} & \cellcolor{basecellcol!32} \textcolor{black}{3} & & \cellcolor{basecellcol!0}\textcolor{black}{0.0\%}\\
\texttt{mistral-large} & \cellcolor{basecellcol!80} \textcolor{black}{6} & \cellcolor{basecellcol!80} \textcolor{black}{6} & \cellcolor{basecellcol!32} \textcolor{black}{3} & \cellcolor{basecellcol!64} \textcolor{black}{5} & \cellcolor{basecellcol!64} \textcolor{black}{5} & \cellcolor{basecellcol!80} \textcolor{black}{6} & & \cellcolor{basecellcol!42}\textcolor{black}{50.0\%}\\
\texttt{olmo2} & \cellcolor{basecellcol!32} \textcolor{black}{3} & \cellcolor{basecellcol!32} \textcolor{black}{3} & \cellcolor{basecellcol!32} \textcolor{black}{3} & \cellcolor{basecellcol!32} \textcolor{black}{3} & \cellcolor{basecellcol!32} \textcolor{black}{3} & \cellcolor{basecellcol!16} \textcolor{black}{2} & & \cellcolor{basecellcol!0}\textcolor{black}{0.0\%}\\
\texttt{phi4} & \cellcolor{basecellcol!48} \textcolor{black}{4} & \cellcolor{basecellcol!48} \textcolor{black}{4} & \cellcolor{basecellcol!48} \textcolor{black}{4} & \cellcolor{basecellcol!48} \textcolor{black}{4} & \cellcolor{basecellcol!48} \textcolor{black}{4} & \cellcolor{basecellcol!48} \textcolor{black}{4} & & \cellcolor{basecellcol!0}\textcolor{black}{0.0\%}\\
\texttt{qwen2.5-coder} & \cellcolor{basecellcol!48} \textcolor{black}{4} & \cellcolor{basecellcol!64} \textcolor{black}{5} & \cellcolor{basecellcol!48} \textcolor{black}{4} & \cellcolor{basecellcol!64} \textcolor{black}{5} & \cellcolor{basecellcol!64} \textcolor{black}{5} & \cellcolor{basecellcol!64} \textcolor{black}{5} & & \cellcolor{basecellcol!0}\textcolor{black}{0.0\%}\\

\bottomrule
\end{tabular}
\end{table}

In contrast, none of the locally executed open weight models achieved a fully validated implementation. Nevertheless, Qwen2.5-Coder generally reached later stages of execution and validation than most other local models, and in several trials progressed further than GPT-4o. Gemma~3 and Phi-4 also frequently reached intermediate stages, i.e., executing without errors or timeouts, indicating that while smaller models can often generate runnable code, successful end-to-end reproduction of PPHPC dynamics remained out of reach for these smaller models.

\subsubsection{Statistical validation against the NetLogo baseline}
\label{sec:results:execvalid:stats}

Table~\ref{tab:pvalues} shows the corrected $p$-values from the Energy test comparing simulation outputs from each LLM-generated implementation against the baseline NetLogo model, evaluated for two parameter sets. Where the $p$-values for the two parameter sets are not significant, the model-trial combination is assigned a score of 6, otherwise it is scored with 5.  Logically, only model-trial combinations that reached at least the statistical comparison stage (i.e., score $\ge 5$) are reported in this analysis; trials that failed earlier in the execution pipeline are excluded.

\begin{table}[!tb]
\centering
\footnotesize

\caption{$p$-values from the Energy test (corrected with the Benjamini-Hochberg procedure) used to classify LLM-generated PPHPC implementations as producing statistically incorrect (score~5) or statistically indistinguishable (success; score~6) simulation output when compared against a baseline NetLogo implementation. For each model and trial/seed, both the LLM-generated implementation and the NetLogo baseline are executed 30 times using different random seeds, yielding 60 condensed outputs (30 per implementation) used in the statistical comparison. The Energy test is applied to the minimum number of principal components required to explain at least 80\% of the variance in these outputs, evaluated under two parameter sets (column ``Param.''). Shaded cells indicate statistically significant differences at the corrected threshold $p<0.01$; non-significant results ($p\ge 0.01$) are treated as statistically indistinguishable (score~6 only if $p\ge 0.01$ for both parameter sets; otherwise score~5). Values below $0.001$ are reported as $<0.001$. Entries marked ``--'' indicate trials for which the test could not be performed due to earlier stage failures.}

\label{tab:pvalues}
\begin{tabular}{lcrrrrrr}
\toprule
 & & \multicolumn{5}{l}{Trial/seed} \\
\cmidrule(){3-8}
Model & Param. & \multicolumn{1}{r}{1/0088} & \multicolumn{1}{r}{2/0288} & \multicolumn{1}{r}{3/3259} & \multicolumn{1}{r}{4/5613} & \multicolumn{1}{r}{5/8153} & \multicolumn{1}{r}{6/8797} \\
\midrule

\texttt{claude-3.7-sonnet} & 1 & \cellcolor{black!10} $<0.001$ & \cellcolor{black!10} $<0.001$ & $0.148$ & $0.676$ & $0.565$ & $0.194$\\
 & 2 & \cellcolor{black!10} $<0.001$ & \cellcolor{black!10} $<0.001$ & $1.000$ & $0.489$ & $0.289$ & $0.148$\\
\addlinespace
\texttt{deepseek-r1} & 1 & $0.157$ & -- & -- & -- & $1.000$ & $0.254$\\
 & 2 & $0.565$ & -- & -- & -- & $0.214$ & \cellcolor{black!10} $<0.001$\\
\addlinespace
\texttt{deepseek-v3} & 1 & \cellcolor{black!10} $<0.001$ & \cellcolor{black!10} $<0.001$ & \cellcolor{black!10} $<0.001$ & -- & $0.441$ & $0.628$\\
 & 2 & \cellcolor{black!10} $<0.001$ & \cellcolor{black!10} $<0.001$ & \cellcolor{black!10} $<0.001$ & -- & $1.000$ & $0.949$\\
\addlinespace
\texttt{gpt-4.1} & 1 & $0.180$ & $0.432$ & $0.154$ & $1.000$ & $0.441$ & $0.083$\\
 & 2 & $0.623$ & $0.930$ & $0.344$ & $0.033$ & $1.000$ & $0.489$\\
\addlinespace
\texttt{grok-3-beta} & 1 & $0.322$ & $0.364$ & \cellcolor{black!10} $<0.001$ & \cellcolor{black!10} $<0.001$ & \cellcolor{black!10} $<0.001$ & $0.206$\\
 & 2 & $1.000$ & $1.000$ & \cellcolor{black!10} $<0.001$ & \cellcolor{black!10} $<0.001$ & \cellcolor{black!10} $<0.001$ & $1.000$\\
\addlinespace
\texttt{mistral-large} & 1 & $0.134$ & $0.501$ & -- & $0.344$ & $0.788$ & $0.064$\\
 & 2 & $0.037$ & $0.013$ & -- & \cellcolor{black!10} $<0.001$ & \cellcolor{black!10} $<0.001$ & $0.148$\\
\addlinespace
\texttt{qwen2.5-coder} & 1 & -- & \cellcolor{black!10} $<0.001$ & -- & \cellcolor{black!10} $<0.001$ & \cellcolor{black!10} $<0.001$ & \cellcolor{black!10} $<0.001$\\
 & 2 & -- & \cellcolor{black!10} $<0.001$ & -- & \cellcolor{black!10} $<0.001$ & \cellcolor{black!10} $<0.001$ & \cellcolor{black!10} $<0.001$\\

\bottomrule
\end{tabular}
\end{table}

Although a corrected significance threshold of $\alpha = 0.01$ was used for classification, the results are typically far from borderline: when an LLM-generated implementation diverges from the baseline, the corresponding $p$-values are consistently very small ($<0.001$). This indicates a largely binary outcome in practice, where generated implementations either closely reproduce the baseline dynamics under the tested conditions or produce evidently different output distributions.

A small number of asymmetric cases illustrate the value of multi-parameter validation. In particular, DeepSeek-R1 and Mistral Large exhibit trials for which parameter set~1 yields statistically indistinguishable output while parameter set~2 reveals significant discrepancies, suggesting that the latter (and more demanding) parameterization increases sensitivity to small implementation errors that may remain undetected under lighter regimes (e.g., due to increased agent interactions). Nonetheless, although agreement under both parameter sets increases confidence that a generated implementation reproduces the baseline, it does not constitute a proof of equivalence over the full parameter space supported by PPHPC. Extending validation to other parameter sets or higher-dimensional configurations of parameter sets 1 and 2 would further strengthen this assessment. The latter case is rendered impractical by the high computational costs of some successful implementations, as will be discussed in Section~\ref{sec:results:runtimeperf}. Additional details are provided in Fig.~\ref{fig:micomp_v1} and Fig.~\ref{fig:micomp_v2} (\ref{app:pcaplots}), which denote the number of principal components used in the Energy test (explaining at least 80\% of the variance) and depict the first two PCA score dimensions, illustrating that statistically significant cases correspond to clearly separated output clusters.

\subsection{Runtime performance of validated implementations}
\label{sec:results:runtimeperf}

\begin{figure}[!p]
    \centering
    \includegraphics[width=1\textwidth]{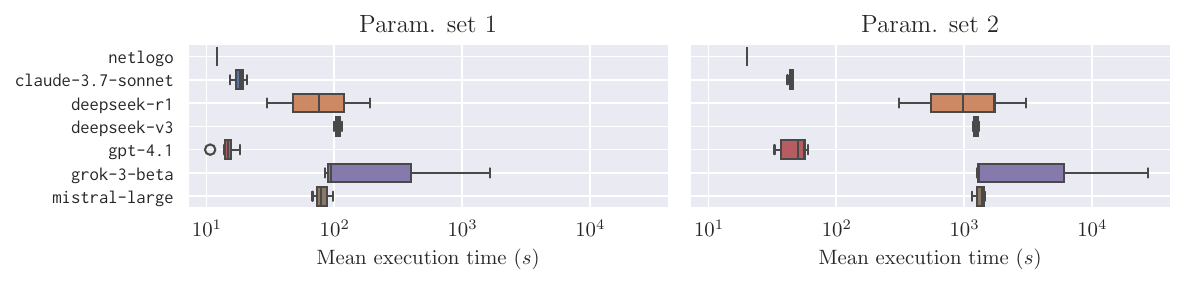}
    \caption{Distribution of mean execution times (seconds) for all successful (score~6) LLM-generated PPHPC implementations over trials/seeds, shown separately for parameter sets 1 and 2. Each data point corresponds to the mean runtime $\bar{t}$ of 30 stochastic replications for a single model-trial/seed combination (as reported in Table~\ref{tab:times}); thus, the box plots summarize a distribution of means rather than individual run times. The NetLogo baseline is included for reference. The logarithmic $x$-axis highlights variation in computational cost both between models and over successful trials for the same model, indicating whether LLMs tend to produce implementations with highly divergent efficiency.}
    \label{fig:meantimes}
\end{figure}

Table~\ref{tab:times} and Fig.~\ref{fig:meantimes} summarize the runtime performance of all statistically validated implementations (score~6), reporting mean execution time over 30 stochastic replications under two parameter sets and normalizing results relative to the NetLogo baseline. Across models, a consistent pattern emerges: parameter set~2 induces a substantially larger computational burden than parameter set~1, and this increase is disproportionately amplified in several LLM-generated implementations. The spread between successful trials is also non-negligible for some models (Fig.~\ref{fig:meantimes}), indicating that implementations that are statistically indistinguishable in output may nonetheless differ substantially in computational efficiency.

\begin{table}[!p]
\centering
\footnotesize

\caption{Mean execution time $\bar{t}$ (seconds) and relative standard deviation $s_\text{rel}$ (percentage) over 30 stochastic replications for all successful LLM-generated PPHPC implementations (score~6), compared against the NetLogo baseline. Column $\times$NL shows the ratio of mean execution time to the NetLogo baseline (values $>1$ indicate slower execution than NetLogo). $s_\text{rel}$ is computed as $100\,s/\bar{t}$, where $s$ is the sample standard deviation of execution time across the 30 runs. Results shown separately for the two parameter sets.}

\label{tab:times}

\begin{tabular}{lrrrrrrr}
\toprule
 &  & \multicolumn{3}{l}{Param. set 1} & \multicolumn{3}{l}{Param. set 2} \\
 \cmidrule(l){3-5}\cmidrule(l){6-8}
 \multicolumn{1}{l}{Model} & \multicolumn{1}{l}{Trial/seed}  & \multicolumn{1}{l}{$\bar{t}\;(\text{s})$} & \multicolumn{1}{l}{$s_\text{rel}\;(\text{\%})$} & \multicolumn{1}{l}{$\times$NL} & \multicolumn{1}{l}{$\bar{t}\;(\text{s})$} & \multicolumn{1}{l}{$s_\text{rel}\;(\text{\%})$} & \multicolumn{1}{l}{$\times$NL} \\
 \midrule

\texttt{netlogo} & -- & 12.13 & 1.35 & 1.00 & 19.84 & 1.70 & 1.00 \\
\cdashline{1-8}[1pt/1pt]
\multirow[t]{4}{*}{\texttt{claude-3.7-sonnet}} & 3/3259 & 15.24 & 1.04 & 1.26 & 41.46 & 1.74 & 2.09 \\
 & 4/5613 & 20.92 & 0.93 & 1.72 & 46.04 & 1.45 & 2.32 \\
 & 5/8153 & 17.84 & 0.92 & 1.47 & 43.59 & 1.29 & 2.20 \\
 & 6/8797 & 18.56 & 1.54 & 1.53 & 45.55 & 1.80 & 2.30 \\
\cdashline{1-8}[1pt/1pt]
\multirow[t]{2}{*}{\texttt{deepseek-r1}} & 1/0088 & 191.14 & 1.95 & 15.76 & 3075.43 & 3.73 & 155.00 \\
 & 5/8153 & 29.86 & 1.45 & 2.46 & 307.64 & 3.55 & 15.50 \\
\cdashline{1-8}[1pt/1pt]
\multirow[t]{2}{*}{\texttt{deepseek-v3}} & 5/8153 & 99.71 & 2.15 & 8.22 & 1171.59 & 1.93 & 59.05 \\
 & 6/8797 & 114.26 & 2.03 & 9.42 & 1315.46 & 2.31 & 66.30 \\
\cdashline{1-8}[1pt/1pt]
\multirow[t]{6}{*}{\texttt{gpt-4.1}} & 1/0088 & 14.66 & 1.00 & 1.21 & 60.20 & 2.87 & 3.03 \\
 & 2/0288 & 13.62 & 2.76 & 1.12 & 33.71 & 1.82 & 1.70 \\
 & 3/3259 & 14.72 & 1.34 & 1.21 & 48.65 & 2.81 & 2.45 \\
 & 4/5613 & 15.81 & 1.87 & 1.30 & 51.95 & 2.28 & 2.62 \\
 & 5/8153 & 10.68 & 1.06 & 0.88 & 32.80 & 1.74 & 1.65 \\
 & 6/8797 & 18.17 & 1.16 & 1.50 & 57.93 & 1.82 & 2.92 \\
\cdashline{1-8}[1pt/1pt]
\multirow[t]{3}{*}{\texttt{grok-3-beta}} & 1/0088 & 84.64 & 1.13 & 6.98 & 1254.45 & 2.61 & 63.22 \\
 & 2/0288 & 1659.46 & 2.62 & 136.80 & 27590.92 & 6.36 & 1390.53 \\
 & 6/8797 & 94.39 & 1.05 & 7.78 & 1308.10 & 1.85 & 65.93 \\
\cdashline{1-8}[1pt/1pt]
\multirow[t]{3}{*}{\texttt{mistral-large}} & 1/0088 & 67.52 & 2.11 & 5.57 & 1154.28 & 2.94 & 58.17 \\
 & 2/0288 & 98.48 & 1.25 & 8.12 & 1449.56 & 1.33 & 73.05 \\
 & 6/8797 & 79.39 & 2.51 & 6.54 & 1387.16 & 2.24 & 69.91 \\

\bottomrule
\end{tabular}
\end{table}

Among the evaluated models, GPT-4.1 produced the most efficient implementations overall, with runtime close to the NetLogo baseline and relatively limited between-trial dispersion. One GPT-4.1 implementation marginally outperformed NetLogo (NL) under parameter set~1 ($0.88\times$NL), while degrading slightly to $1.65\times$NL under parameter set~2. Interestingly, the runtime performance of GPT-4.1-generated models does not scale proportionally between parameter sets. Although the fastest implementation remains the fastest under both configurations, the relative ordering of slower trials changes, indicating that increased computational load stresses different code paths unevenly rather than inducing a uniform slowdown. Claude~3.7 Sonnet likewise generated consistently efficient implementations when successful, remaining within $1.26$--$1.72\times$NL for parameter set~1 and exhibiting particularly tight variation for parameter set~2 ($2.09$--$2.32\times$NL), with relative performance ordering preserved between parameter sets.

In contrast, the remaining validated models displayed substantially higher runtimes and, in some cases, extreme inefficiency under parameter set~2. DeepSeek-V3 produced relatively consistent but slow implementations (approximately $8$--$9\times$NL for parameter set~1 and $59$--$66\times$NL for parameter set~2), whereas DeepSeek-R1 displayed noticeably higher variability, ranging from moderately slow to severely inefficient depending on the trial. Grok~3 and Mistral Large further illustrate that successful replication of baseline output does not imply computational competitiveness: both produced implementations that were orders of magnitude slower than NetLogo under parameter set~2, including one Grok trial exceeding $10^3\times$NL. Overall, these results highlight that runtime efficiency is a distinct axis of performance from statistical correctness, and that validated LLM-generated implementations can differ substantially in computational cost even when they reproduce comparable simulation outputs.

\subsection{Static code quality of validated implementation}
\label{sec:results:codequality}

\begin{figure}
    \centering
    \includegraphics[width=0.9\textwidth]{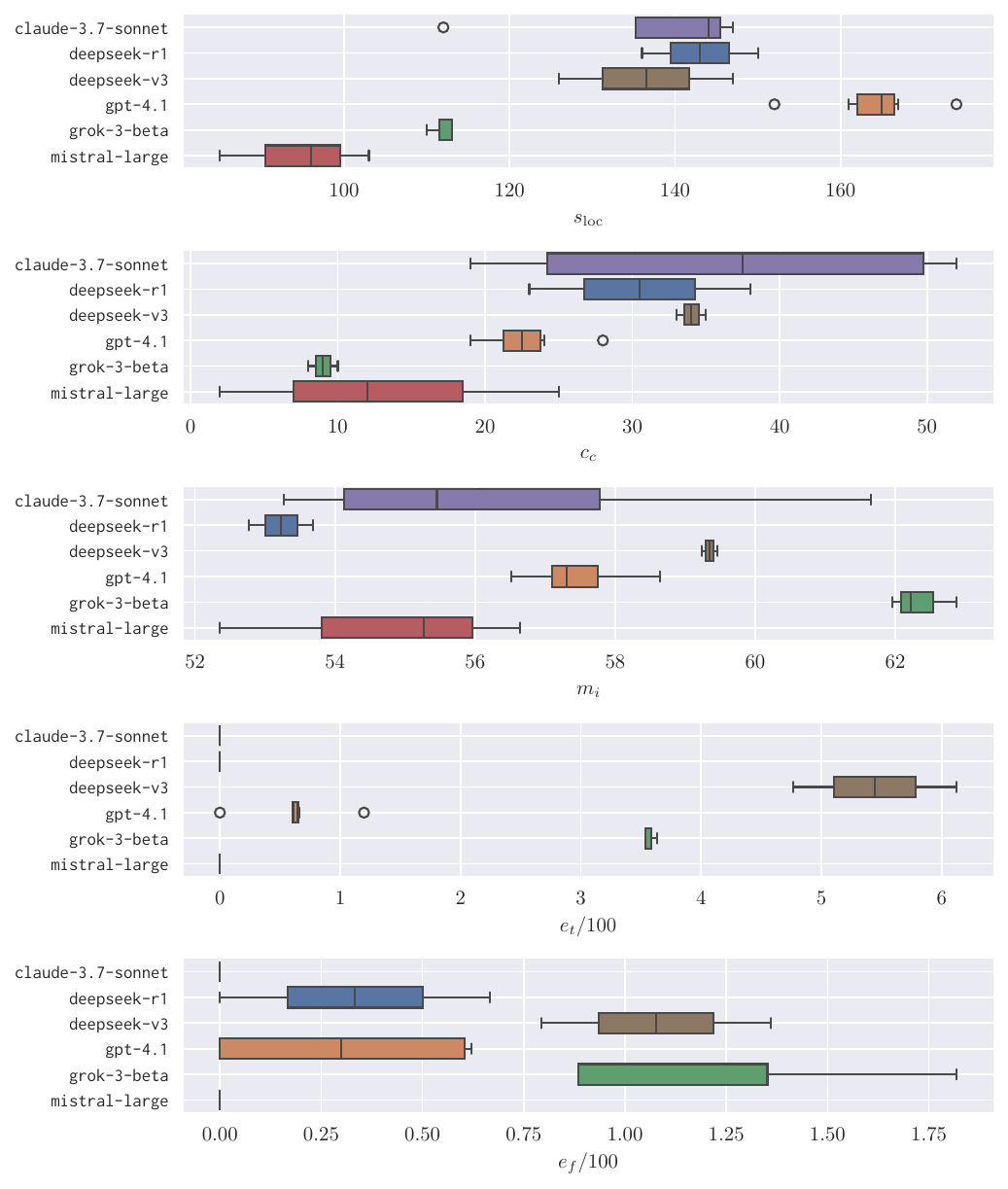}
    \caption{Distribution of code quality metrics for all successful (score~6) LLM-generated PPHPC implementations over trials/seeds, shown as box plots for each model. Panels report (top to bottom) $s_\mathrm{loc}$ (source lines of code), $c_c$ (cyclomatic complexity; lower is better), $m_i$ (maintainability index; 0--100; higher is better), $e_t/100$ (type warnings per 100~$s_\mathrm{loc}$), and $e_F/100$ (flaws and formatting warnings per 100~$s_\mathrm{loc}$ indicating potential code quality issues). This figure complements Table~\ref{tab:metrics} by displaying between-trial variability in code structure and static analysis warnings, highlighting whether a given LLM tends to produce consistently similar implementations or highly divergent code among its successful trials.}
    \label{fig:codequality}
\end{figure}

Table~\ref{tab:metrics} and Fig.~\ref{fig:codequality} summarize static code quality metrics for all statistically validated implementations (score~6), aggregated by model over successful trials. Median source lines of code varied substantially across models, from relatively compact implementations produced by Mistral Large ($s_\mathrm{loc}=96$) and Grok~3 ($s_\mathrm{loc}=113$) to more verbose implementations from GPT-4.1 ($s_\mathrm{loc}=165$) and Claude~3.7 Sonnet ($s_\mathrm{loc}=144$). Cyclomatic complexity likewise differed considerably: Grok yielded the lowest median $c_c$ (9), whereas Claude exhibited the highest median $c_c$ (37.5) and the widest between-trial dispersion (Fig.~\ref{fig:codequality}), implying greater structural variability among its successful implementations.

\begin{table}[!tb]
\centering
\footnotesize
\caption{Median code quality metrics for successful (score~6) LLM-generated PPHPC implementations, aggregated by model over trials/seeds. Column $n$ reports the total number of successful instances for each model. For each metric, values shown are the median over these $n$ successful implementations: $s_\mathrm{loc}$ denotes source lines of code (non-comment lines of code); $c_c$ is cyclomatic complexity (number of independent code paths; lower is better); $m_i$ is the maintainability index (0--100; higher is better); $e_t/100$ is the number of type warnings per 100~$s_\mathrm{loc}$; and $e_F/100$ is the number of flaws and formatting warnings per 100~$s_\mathrm{loc}$, reflecting potential code quality issues flagged by static analysis.}
\label{tab:metrics}

\begin{tabular}{lrrrrrr}
\toprule
Model & $n$ & $s_\mathrm{loc}$ & $c_c$ & $m_i$ & $e_t/100$ & $e_f/100$ \\
\midrule
\texttt{claude-3.7-sonnet} & 4 & 144.00 & 37.50 & 55.45 & 0.00 & 0.00 \\
\texttt{deepseek-r1} & 2 & 143.00 & 30.50 & 53.23 & 0.00 & 0.33 \\
\texttt{deepseek-v3} & 2 & 136.50 & 34.00 & 59.34 & 5.44 & 1.08 \\
\texttt{gpt-4.1} & 6 & 165.00 & 22.50 & 57.30 & 0.61 & 0.30 \\
\texttt{grok-3-beta} & 3 & 113.00 & 9.00 & 62.22 & 3.54 & 0.88 \\
\texttt{mistral-large} & 3 & 96.00 & 12.00 & 55.27 & 0.00 & 0.00 \\
\bottomrule
\end{tabular}
\end{table}

Maintainability index ($m_i$) and static analysis warning rates further emphasize that these metrics capture distinct facets of code quality. Grok and DeepSeek-V3 achieved the highest median $m_i$ values (62.22 and 59.34, respectively), despite exhibiting comparatively elevated type and flaws/formatting issue rates per 100~$s_\mathrm{loc}$ (especially for DeepSeek-V3: $e_t/100=5.44$ and $e_f/100=1.08$). In contrast, Claude and Mistral produced error-free implementations under these checks, despite highly variant $m_i$ values. GPT-4.1, while strongest in end-to-end success and efficiency, did not dominate any single static metric, combining the highest median $s_\mathrm{loc}$ with intermediate $c_c$ and $m_i$ and non-zero warning rates. Overall, the distributions in Fig.~\ref{fig:codequality} indicate that even among validated implementations, models differ both in typical code structure and in the consistency with which they reproduce similar static profiles between trials.

\section{Discussion}
\label{sec:discussion}

This study examined whether contemporary LLMs can (i) consistently generate executable Python implementations of an agent-based model from a complete ODD specification that are statistically indistinguishable from a validated baseline (RQ1), and (ii) how the quality of statistically validated LLM-generated implementations varies between models and over successful trials of the same model (RQ2), considering both runtime efficiency and maintainability-relevant code properties.

\subsection*{RQ1: Reliability of LLMs in generating statistically valid ABM implementations}

The results show that end-to-end generation of statistically validated ABM implementations from an ODD description is feasible for several contemporary hosted models, but reliability is strongly model-dependent. In this setting, ``reliability'' is reflected by the probability of reaching score~6 over stochastic trials (Table~\ref{tab:results}; Fig.~\ref{fig:scores}), where different trials reflect variation arising from the models' generative sampling. Hosted models that repeatedly achieved success did so by consistently producing complete implementations that simultaneously satisfied three conditions: (i) syntactic correctness, (ii) successful execution with the expected output interface, and (iii) output dynamics that remained statistically indistinguishable from the baseline under multiple parameterizations. Conversely, models that failed to reach score~6 typically did not fail at code production per se, but rather exhibited failures at intermediate stages---including runtime errors, incorrect output formatting, or statistically significant deviations in simulation output---indicating that producing runnable code is substantially easier than reproducing correct model dynamics.

GPT-4.1 was the only model to achieve score~6 in all trials, indicating the highest robustness under the experimental protocol. Claude~3.7 Sonnet constituted a close second, achieving valid implementations in 4 out of 6 trials, while other hosted models (Mistral Large, Grok~3, DeepSeek-R1, DeepSeek-V3) succeeded only intermittently. Thus, although several contemporary hosted models were capable of producing statistically valid implementations at least occasionally, reliability varied substantially between models.

From a modeling perspective, this variability is consequential because ABMs are highly implementation-sensitive, and partial failures (e.g., producing code that runs but yields statistically different output) may be difficult to detect without formal output validation. In this sense, the PPHPC setting highlights that correctness in ABM code generation is not adequately captured by the ability to produce executable code, but depends on whether the generated implementation reproduces the expected output distributions. This finding empirically substantiates recent calls to move beyond surface-level code generation toward behaviorally grounded evaluation of AI-assisted modeling workflows, as emphasized by both \citet{andelfinger2025intelligent} and \citet{berger2024towards}.

The lack of success among locally executed open weight models further reinforces this distinction. Several local models produced runnable code and reached intermediate pipeline stages, but none produced statistically validated implementations. This contrasts with prior evidence that models such as Phi-4 and LLaMA-3.3 can perform well in code generation for context-specific engineering tasks~\citep{fernandes2025deepseek}. In the present context, the bottleneck appears not to be basic code writing, but rather the precise translation of an ABM specification into an implementation that preserves stochastic dynamics under validation. This suggests that ABM replication from ODD descriptions places comparatively higher demands on correctly implementing interaction rules, stochastic mechanisms, and data collection pipelines than many conventional programming tasks, even when the underlying model is conceptually simple and well described.

\subsection*{RQ2: Variation in runtime efficiency and maintainability among statistically valid implementations}

With respect to statistically validated implementations, the results show that code quality is not uniform among LLMs and cannot be inferred from correctness alone. GPT-4.1 produced implementations that were consistently efficient over trials, with relatively limited dispersion in runtime performance compared to other models (Table~\ref{tab:times}; Fig.~\ref{fig:meantimes}). In parallel, its successful implementations displayed broadly favorable maintainability-relevant properties (Table~\ref{tab:metrics}; Fig.~\ref{fig:codequality}), indicating that GPT-4.1 not only succeeds reliably, but also tends to generate implementations whose computational cost and static code properties remain relatively stable between stochastic trials.

Claude~3.7 Sonnet exhibited a complementary profile. Although it was less reliable than GPT-4.1 in attaining statistical validation (4 out of 6 successful trials), its successful implementations showed consistent and competitive runtimes throughout trials, particularly under the more demanding parameter set~2 (Table~\ref{tab:times}). At the code level, however, Claude's validated implementations had the highest median cyclomatic complexity and substantial variability over successful trials (Fig.~\ref{fig:codequality}), indicating that efficiency gains can co-occur with more complex control structure. This separation between runtime behavior and structural simplicity shows that efficiency and maintainability are partially decoupled quality dimensions in LLM-generated implementations.

Conversely, DeepSeek-R1, DeepSeek-V3, Grok~3, and Mistral Large demonstrate that validated correctness can coexist with highly variable and sometimes extreme computational cost. Several of these models produced implementations that were orders of magnitude slower than others under parameter set~2 (Fig.~\ref{fig:meantimes}), despite remaining statistically indistinguishable in output for the tested conditions. In addition, the static analysis results indicate that higher maintainability index values do not necessarily correspond to fewer warnings, reinforcing the value of reporting multiple quality metrics rather than relying on any single proxy for maintainability.

\subsection*{Implications}

Overall, these findings support two broader implications. First, statistical validation should be treated as a prerequisite but not a sufficient condition for adoption of LLM-generated ABM code: runtime efficiency and maintainability properties remain highly model- and trial-dependent among validated cases. This distinction is particularly important in light of emerging ``vibe coding'' practices, i.e., relying on LLM-generated code mainly because it appears to work, without carefully inspecting or fully understanding the implementation, thereby lowering the barrier to deploying inefficient or fragile implementations~\citep{ge2025survey}. Second, the strongest models in correctness (GPT-4.1, followed by Claude) also tended to generate higher-quality implementations in aggregate, but the remaining validated models reveal that the same output-level behavior may be produced by implementations with widely different computational and structural characteristics. In practice, this motivates evaluation protocols for LLM-generated ABM code that explicitly separate \emph{correctness} from \emph{quality}, and treat both as necessary for dependable use in simulation research.

More generally, the appropriateness of LLM-generated ABM code depends on the purpose for which the model is being built~\citep{edmonds2019different}. For replication-oriented tasks such as the one studied here, output level statistical validation is a natural first requirement. Conversely, if the ABM is intended to support explanation in terms of mechanisms, then additional checks are needed to determine whether the generated code preserves the intended causal structure, scheduling logic, and local interaction rules in a transparent way. Likewise, for teaching-oriented uses, a simpler and more readable implementation might be preferable even if a more compact or faster alternative reproduces the same outputs. Thus, the present results should be interpreted as evidence about one important dimension of adequacy, i.e., behavioral replication, not as a complete substitute for purpose-dependent evaluation of generated ABM code.

\section{Limitations}
\label{sec:lims}

First, this work is a single-model case study centered on PPHPC, a well-documented and successfully replicated ABM. Although this enables strong verification of implementation correctness, it limits generalizability to ABMs with different characteristics (e.g., dissimilar spatial structure, network interactions, continuous time dynamics, or more complex agent cognition). Moreover, because PPHPC and its ODD specification have been openly available since 2015, and since reference implementations exist in multiple languages (including NetLogo, Java, and C/OpenCL), it is possible that some of the evaluated LLMs were exposed during training to descriptions of the model or to implementations in other programming languages, introducing a potential data leakage effect. However, to the best of our knowledge no publicly available Python implementation of PPHPC exists, thus successful generations cannot rely on direct memorization or translation of a Python reference implementation. Eliminating exposure risk entirely would nonetheless require the development of a new, previously unseen ABM specifically for this study, which would entail substantial effort in model design, specification, and validation, and would forgo the advantages of using a well-studied reference model with established statistical behavior, replication history, and prior discussion of implementation reuse challenges~\citep{berger2024towards}. Finally, published ABM descriptions are frequently incomplete or ambiguous, which is a known barrier to replication and reuse~\citep{donkin2017replicating,david2017vvs,grimm2025using}; in this sense, PPHPC and its detailed ODD description likely represent an upper-bound scenario for specification quality, and the present findings should be interpreted accordingly when considering less formal or less complete model documentation.

Second, due to practical cost and technical constraints, the evaluation relied on a limited number of trials (six generations per model) and was performed under a single fixed prompt rather than a broader set of prompt formulations. While this design is sufficient to highlight substantial cross-model differences and within-model variability, which is the primary objective of the study, it does not support precise estimation of success probabilities or a strong separation between systematic failure modes and sampling noise. Nonetheless, the sampling temperature ($T=0.1$) allows for a small degree of variability, partially mitigating sensitivity to sampling effects and to the absence of systematic testing across multiple prompt variants. Moreover, prior work indicates that at least some degree of variability is intrinsic rather than purely prompt-induced~\citep{rajput2025dynamic}.

Third, statistical validation was performed under only two parameter sets. While this is more informative than single configuration testing, it may still fail to expose implementation errors that only manifest under other regions of the parameter space (e.g., extreme densities, or edge conditions). More generally, statistical indistinguishability 
should be interpreted as evidence of agreement under the tested conditions rather than a proof of full behavioral equivalence over all observables, time horizons, or parameterizations.

Fourth, no expert-crafted Python baseline was employed for absolute performance and code quality benchmarking. Runtime comparisons against the NetLogo reference therefore provide only coarse context, since NetLogo operates under a different language and runtime (Java-based) and is not necessarily optimized for raw execution speed. This matters because alternative validated implementations of PPHPC demonstrate that substantially higher performance is achievable: a single-threaded Java implementation can exceed NetLogo performance by more than $6\times$, while parallel and low-level implementations may be one to two orders of magnitude faster~\citep{fachada2015parallelization,fachada2017assessing}. Consequently, performance, as well as static code quality, should be interpreted relatively between LLM-generated Python implementations, rather than against an ``optimal'' reference.

Fifth, static analysis metrics provide only partial proxies for maintainability. Measures such as cyclomatic complexity, maintainability index, and warning counts correlate with maintenance effort, but they do not fully capture readability, modularity, testability, or ease of extension, and they may miss defects that emerge during iterative modification. Furthermore, no human evaluation of code understandability or editability was performed; the analysis therefore reflects automated, scalable assessment practices commonly used in prior evaluations of LLM-generated code, but cannot confirm how safely practitioners could adapt the generated implementations in realistic development settings~\citep{simoes2024evaluating,wang2025maintaincoder}. 

Finally, the pace of LLM development implies that results obtained under a fixed experimental window rapidly become a historical snapshot. The presented findings should therefore be interpreted as characterizing the state of evaluated models and tooling as of July~2025, rather than as a permanent ranking of LLM capabilities for ABM implementation from ODD descriptions. Nonetheless, despite these constraints, the study provides a grounded, multi-criterion view of what can currently be expected from LLMs in this task, including success rates, validation outcomes, and the variability of performance and maintainability-relevant properties across models and trials.

\section{Conclusions}
\label{sec:conclusions}

This study examined whether contemporary large language models can translate a complete, standardized agent-based model specification into executable implementations that are not only runnable, but also behaviorally faithful and practically usable. Using the PPHPC benchmark and its ODD specification as a controlled test case, we evaluated a heterogeneous set of modern LLMs under a fixed prompting and execution protocol, combining staged execution outcomes, model-independent statistical output validation, runtime measurements, and static code quality analysis.

The results show that specification-to-code generation for ABMs is now feasible in principle: a small subset of evaluated LLMs was able to produce Python implementations whose stochastic output was statistically indistinguishable from a validated NetLogo baseline. At the same time, success was strongly model-dependent and probabilistic, with many generated implementations failing at intermediate stages or producing simulations that were runnable yet behaviorally incorrect. These findings reinforce that executability alone is an insufficient criterion for assessing LLM-generated simulation code, particularly for ABMs where small implementation differences can materially affect emergent dynamics.

Crucially, even among statistically validated implementations, substantial variation was observed in runtime efficiency and maintainability-relevant code properties. Some implementations exhibited stable and competitive performance with relatively few structural issues, while others suffered from severe performance degradation or increased structural complexity despite producing correct outputs. This decoupling of behavioral correctness from computational and structural quality highlights the need to treat correctness, efficiency, and maintainability as distinct evaluation dimensions when assessing LLM-generated simulation models.

These findings suggest that LLMs can serve as useful aids for implementing agent-based models from standardized textual specifications, but only when embedded within rigorous evaluation workflows that include behavioral validation and quality assessment. For reproducible simulation practice, this implies that LLM-generated models should be treated as candidate implementations rather than authoritative realizations of a specification, and subjected to the same verification, validation, and performance scrutiny as human-written code. In practice, this means that researchers should use LLMs primarily for rapid prototyping or drafting implementations, and only adopt generated code after checking executability, comparing model behavior against a trusted baseline or expected patterns, and assessing whether runtime performance and code structure are adequate for the intended use. Which evaluation workflow is appropriate, however, depends on the purpose of the model. Behavioral similarity may be sufficient for some replication-oriented uses, but mechanism-oriented or teaching-oriented applications may require stronger scrutiny of internal faithfulness, transparency, and code simplicity.

Future work should extend this evaluation framework to additional agent-based models with different structural characteristics, explore the sensitivity of results to alternative prompting and refinement strategies, and investigate how LLM-based code generation might be combined with modular model components or automated testing to improve reliability. As LLM capabilities continue to evolve, systematic, multi-criterion assessments such as the one presented here will remain essential for understanding when and how these tools can be responsibly integrated into simulation modeling workflows.

\section*{Data Availability}

The data generated by this study and its respective analysis are available at \url{https://doi.org/10.5281/zenodo.18521255} under the CC-BY 4.0 (data and outputs) and MIT licenses (code) \citep{fachada2026suppl}.

\section*{Declaration of competing interest}

The authors declare that they have no known competing financial interests or personal relationships that could have appeared to influence the work reported in this paper.

\section*{Declaration of generative AI and AI-assisted technologies in the manuscript preparation process}

During the preparation of this manuscript, the authors used ChatGPT (OpenAI) to support language editing, improve text flow, check acronym expansion on first use, and obtain limited technical clarifications related to \LaTeX{} and Python (e.g., Pandas usage), as well as high-level suggestions on section organization and draft wording for the abstract and conclusions. All scientific ideas, model design choices, experimental rationale, data analysis, interpretation of results, and conclusions were developed by the authors. All content generated with AI assistance was carefully reviewed, edited, and, where appropriate, substantially modified by the authors, who take full responsibility for the final content of the published article.

\section*{Acknowledgements}

This research was partially funded by the Funda\c c\~ao para a Ci\^encia e a Tecnologia (Portugal, \url{https://ror.org/00snfqn58}) under grants/projects 
UID/06486/2025 (\url{https://doi.org/10.54499/UID/06486/2025}), 
UID/PRR/06486/2025 (\url{https://doi.org/10.54499/UID/PRR/06486/2025}), 
UID/PRR2/06486/2025 (\url{https://doi.org/10.54499/UID/PRR2/06486/2025}), 
\linebreak
UID/50008/2025 (\url{https://doi.org/10.54499/UID/50008/2025}), 
UID/PRR/50008/2025 \linebreak (\url{https://doi.org/10.54499/UID/PRR/50008/2025}),
UID/PRR2/50008/2025 (\url{https://doi.org/10.54499/UID/PRR2/50008/2025}),
2023.15441.TENURE.051/CP00003/CT00029, \linebreak 
UID/00408/2025 (\url{https://doi.org/10.54499/UID/00408/2025}), 
UID/PRR/00408/2025 \linebreak (\url{https://doi.org/10.54499/UID/PRR/00408/2025}), 
and, CEECINST/00002/2021/CP2788\-/\-CT0001 (\url{https://doi.org/10.54499/CEECINST/00002/2021/CP2788/CT0001}), 
as well as by the Instituto Lusófono de Investigação e Desenvolvimento (Portugal, \url{https://ror.org/02qy8ba98}) under Project COFAC/ILIND/COPELABS/1/2024, 
and by the Ministerio de Ciencia, Innovación y Universidades (Spain, \url{https://ror.org/05r0vyz12}) -- MICIU/AEI/10.13039\-/\-501100011033 under project number PID2023-147409NB-C21. 

\appendix

\section{Glossary of LLM-related terms}
\label{app:glossary}

This glossary briefly explains relevant LLM- and code quality-related terms used in the paper for readers who may be less familiar with the associated terminology.

\begin{description}

\item[Code-oriented model / code-specialised training]
A language model trained or further tuned with a stronger emphasis on programming tasks, such as code completion, code generation, and code understanding. In contrast, a \emph{general-purpose} model is trained for a broader range of language tasks and is not primarily optimized for software development.

\item[Context window / maximum context window in tokens]
The maximum amount of text a model can consider at one time when generating a response. It is usually measured in \emph{tokens}, which are small units of text roughly corresponding to parts of words, whole words, punctuation, or symbols.

\item[Dense architecture]
A model architecture in which, for a given input, all or most model parameters participate in the computation of each generated token. Dense models contrast with mixture-of-experts architectures, where only part of the model is activated at a time.

\item[Generalist model]
A model intended for broad use across many different tasks, such as question answering, summarization, reasoning, and code generation, rather than being specialized for a single domain.

\item[Hosted model / proprietary hosted access]
A model accessed remotely through a provider's online service or application programming interface (API). The user can query the model, but the model weights and serving infrastructure are not directly available for local execution or inspection.

\item[Instruction following]
The ability of a model to accurately follow the user's prompt and constraints, for example respecting required function names, libraries, output formats, and other task requirements.

\item[Long-context inference / short-context inference]
\emph{Inference} is the process of generating an output from a trained model. \emph{Long-context} inference refers to models that can process larger amounts of input text at once, whereas \emph{short-context} models can consider only smaller inputs before older text must be truncated or omitted.

\item[Mixture-of-experts (MoE) architecture]
A model architecture in which only a subset of specialized internal components (``experts'') is activated for each token or input segment. This can increase effective model capacity without requiring all parameters to be used on every computation.

\item[Open-weight model / deployable open-weight execution]
A model whose trained parameters (weights) are publicly released or otherwise available for local deployment by users. Such models can be executed on the user's own hardware, which may improve transparency and reproducibility relative to hosted-only systems.

\item[Parameter count]
The number of adjustable numerical values in a model. It is commonly used as a rough indicator of model size, although models with similar parameter counts can still differ substantially in capability.

\item[Sampling temperature]
A parameter that influences how variable or random a model's generated text is. Lower temperature values generally make outputs more conservative and repeatable, while higher values usually increase diversity but also unpredictability.

\item[Static code quality]
Properties of source code that can be assessed without running the program, such as code length, branching complexity, maintainability-related metrics, and recurring patterns associated with poor design or harder future maintenance.

\item[Vibe coding]
An informal term for a style of programming in which users rely heavily on LLM-generated code based mainly on whether it seems to work in practice, often without closely inspecting or fully understanding the underlying implementation.

\end{description}

\section{PCA score plots: NetLogo baseline vs. LLM-generated models}
\label{app:pcaplots}

\begin{figure}
    \centering
    \includegraphics[width=1\textwidth]{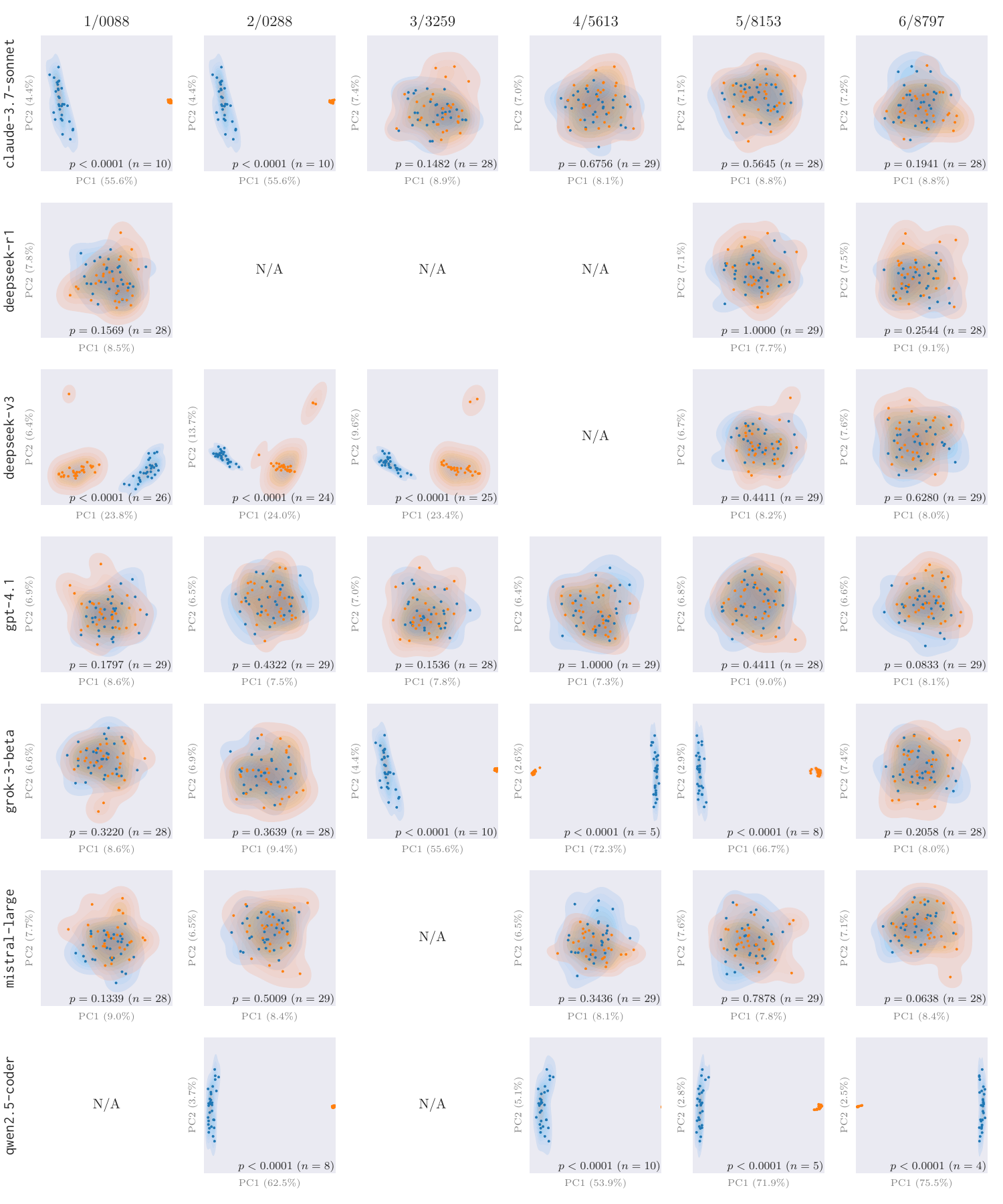}
    \caption{PCA score plots (PC1 vs PC2) and respective distribution density of the baseline NetLogo model runs (in blue) versus the LLM-generated model runs (in orange) for parameter set 1. Rows correspond to different LLM-generated models (labeled at the left), while columns are associated with the trial/seed (labeled at the top) which generated the respective implementation. Percentages in parenthesis after the PC1/PC2 labels correspond to the ratio of explained variance of the corresponding PC. The $p$-value at the bottom left of each figure is drawn from the Energy test applied to the number of PCs ($n$) that explain at least 80\% of the variance.}
    \label{fig:micomp_v1}
\end{figure}

\begin{figure}
    \centering
    \includegraphics[width=1\textwidth]{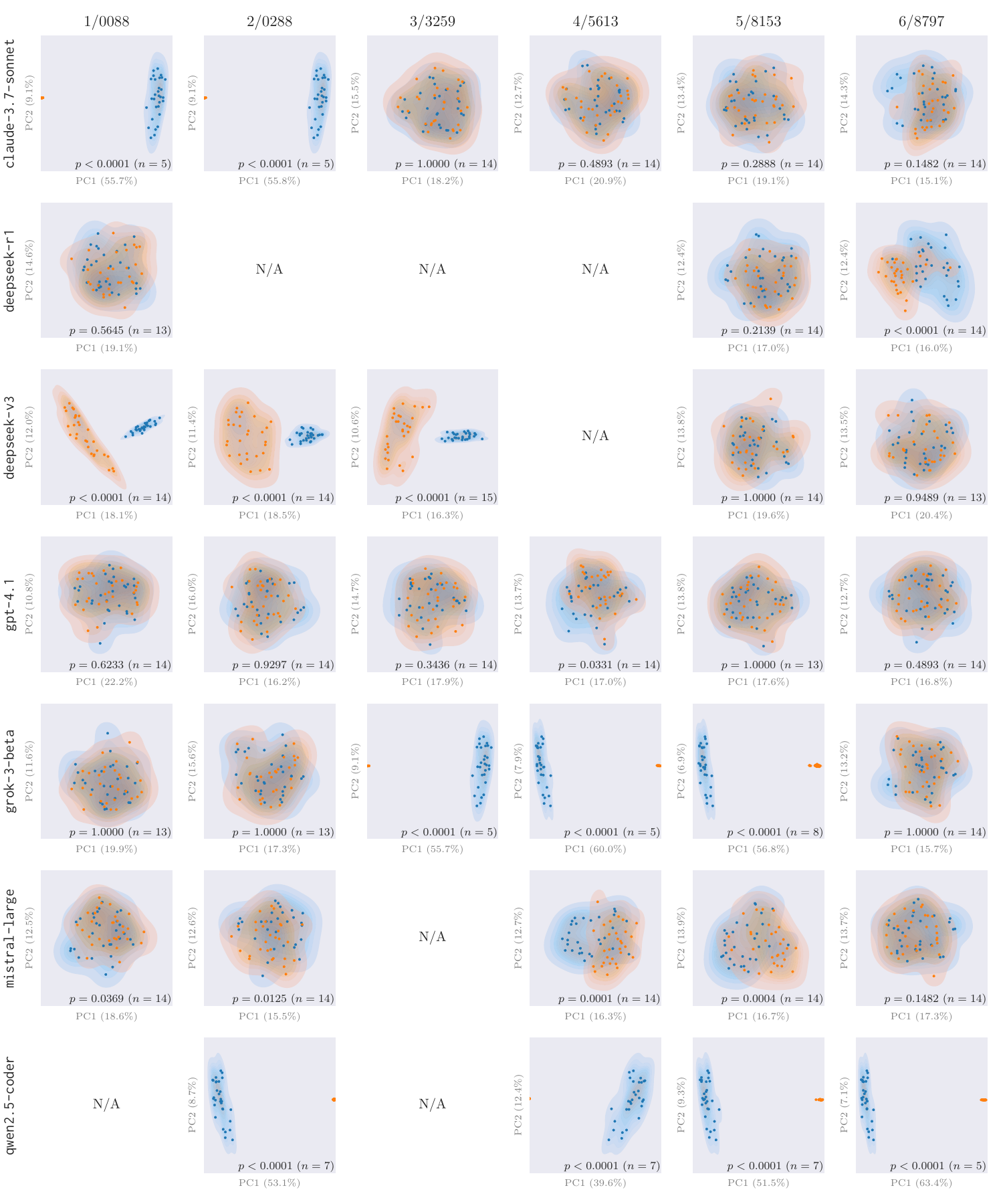}
    \caption{PCA score plots (PC1 vs PC2) and respective distribution density of the baseline NetLogo model runs (in blue) versus the LLM-generated model runs (in orange) for parameter set 2. Rows correspond to different LLM-generated models (labeled at the left), while columns are associated with the trial/seed (labeled at the top) which generated the respective implementation. Percentages in parenthesis after the PC1/PC2 labels correspond to the ratio of explained variance of the corresponding PC. The $p$-value at the bottom left of each figure is drawn from the Energy test applied to the number of PCs ($n$) that explain at least 80\% of the variance.}
    \label{fig:micomp_v2}
\end{figure}

Fig.~\ref{fig:micomp_v1} depicts the PCA score plots (PC1 vs PC2) and the respective distribution density of the baseline NetLogo model runs versus the LLM-generated model runs for parameter set 1. Fig.~\ref{fig:micomp_v2} presents the same information for parameter set 2. The two figures, which complement Table~\ref{tab:pvalues}, also present, at the bottom left of each sub-figure, the $p$-value drawn from the Energy test applied to the number of PCs that explain at least 80\% of the variance ($n$, in parentheses).

\bibliographystyle{elsarticle-harv}

\begin{thebibliography}{78}
\expandafter\ifx\csname natexlab\endcsname\relax\def\natexlab#1{#1}\fi
\providecommand{\url}[1]{\texttt{#1}}
\providecommand{\href}[2]{#2}
\providecommand{\path}[1]{#1}
\providecommand{\DOIprefix}{doi:}
\providecommand{\ArXivprefix}{arXiv:}
\providecommand{\URLprefix}{URL: }
\providecommand{\Pubmedprefix}{pmid:}
\providecommand{\doi}[1]{\href{http://dx.doi.org/#1}{\path{#1}}}
\providecommand{\Pubmed}[1]{\href{pmid:#1}{\path{#1}}}
\providecommand{\bibinfo}[2]{#2}
\ifx\xfnm\relax \def\xfnm[#1]{\unskip,\space#1}\fi
\bibitem[{Abdin et~al.(2024)Abdin, Aneja, Behl, Bubeck, Eldan, Gunasekar,
  Harrison, Hewett, Javaheripi, Kauffmann et~al.}]{abdin2024phi}
\bibinfo{author}{Abdin, M.}, \bibinfo{author}{Aneja, J.},
  \bibinfo{author}{Behl, H.}, \bibinfo{author}{Bubeck, S.},
  \bibinfo{author}{Eldan, R.}, \bibinfo{author}{Gunasekar, S.},
  \bibinfo{author}{Harrison, M.}, \bibinfo{author}{Hewett, R.J.},
  \bibinfo{author}{Javaheripi, M.}, \bibinfo{author}{Kauffmann, P.}, et~al.,
  \bibinfo{year}{2024}.
\newblock \bibinfo{title}{Phi-4 technical report}.
\newblock \bibinfo{journal}{arXiv preprint} ,
  \bibinfo{pages}{2412.08905}\DOIprefix\doi{10.48550/arXiv.2412.08905}.
  \bibinfo{note}{cs.CL}.
\bibitem[{Ahmed et~al.(2025)Ahmed, Azwad and Choudhury}]{ahmed2025simcode}
\bibinfo{author}{Ahmed, T.}, \bibinfo{author}{Azwad, M.M.},
  \bibinfo{author}{Choudhury, S.}, \bibinfo{year}{2025}.
\newblock \bibinfo{title}{{SIMCODE}: A benchmark for natural language to ns-3
  network simulation code generation}, in: \bibinfo{booktitle}{2025 IEEE 50th
  Conference on Local Computer Networks (LCN)}, \bibinfo{publisher}{IEEE}. pp.
  \bibinfo{pages}{1--8}.
\newblock \DOIprefix\doi{10.1109/LCN65610.2025.11146057}.
\bibitem[{Akhavan and Jalali(2024)}]{akhavan2024generative}
\bibinfo{author}{Akhavan, A.}, \bibinfo{author}{Jalali, M.S.},
  \bibinfo{year}{2024}.
\newblock \bibinfo{title}{Generative {AI} and simulation modeling: how should
  you (not) use large language models like {ChatGPT}}.
\newblock \bibinfo{journal}{System Dynamics Review} \bibinfo{volume}{40},
  \bibinfo{pages}{e1773}.
\newblock \DOIprefix\doi{10.1002/sdr.1773}.
\bibitem[{Andelfinger et~al.(2025)Andelfinger, Pellegrini, Carothers, Loper,
  Tan, Wolf and Cai}]{andelfinger2025intelligent}
\bibinfo{author}{Andelfinger, P.}, \bibinfo{author}{Pellegrini, A.},
  \bibinfo{author}{Carothers, C.D.}, \bibinfo{author}{Loper, M.},
  \bibinfo{author}{Tan, W.J.}, \bibinfo{author}{Wolf, V.},
  \bibinfo{author}{Cai, W.}, \bibinfo{year}{2025}.
\newblock \bibinfo{title}{Intelligent modeling and simulation life cycle}.
\newblock \bibinfo{journal}{TechRxiv}
  \DOIprefix\doi{10.36227/techrxiv.174123246.67204848/v2}.
\bibitem[{{Anthropic Team}(2025)}]{anthropic2025claude37}
\bibinfo{author}{{Anthropic Team}}, \bibinfo{year}{2025}.
\newblock \bibinfo{title}{Claude 3.7 Sonnet System Card}.
\newblock \bibinfo{type}{Technical Report}. Anthropic PBC.
\newblock \URLprefix
  \url{https://www.anthropic.com/claude-3-7-sonnet-system-card}.
\bibitem[{{Astral Team}(2022)}]{ruff2022}
\bibinfo{author}{{Astral Team}}, \bibinfo{year}{2022}.
\newblock \bibinfo{title}{Ruff}.
\newblock \bibinfo{howpublished}{\url{https://docs.astral.sh/ruff/}}.
\newblock \bibinfo{note}{Last access: Aug. 24, 2025}.
\bibitem[{Axtell et~al.(1996)Axtell, Axelrod, Epstein and
  Cohen}]{axtell1996aligning}
\bibinfo{author}{Axtell, R.}, \bibinfo{author}{Axelrod, R.},
  \bibinfo{author}{Epstein, J.M.}, \bibinfo{author}{Cohen, M.D.},
  \bibinfo{year}{1996}.
\newblock \bibinfo{title}{Aligning simulation models: A case study and
  results}.
\newblock \bibinfo{journal}{Computational \& Mathematical Organization Theory}
  \bibinfo{volume}{1}, \bibinfo{pages}{123--141}.
\newblock \DOIprefix\doi{10.1007/BF01299065}.
\bibitem[{Benjamini and Hochberg(1995)}]{benjamini1995controlling}
\bibinfo{author}{Benjamini, Y.}, \bibinfo{author}{Hochberg, Y.},
  \bibinfo{year}{1995}.
\newblock \bibinfo{title}{Controlling the false discovery rate: a practical and
  powerful approach to multiple testing}.
\newblock \bibinfo{journal}{Journal of the Royal Statistical Society: Series B
  (Methodological)} \bibinfo{volume}{57}, \bibinfo{pages}{289--300}.
\newblock \DOIprefix\doi{10.1111/j.2517-6161.1995.tb02031.x}.
\bibitem[{Berger et~al.(2024)Berger, Bell, Barton, Chappin, Dre{\ss}ler,
  Filatova, Fronville, Lee, van Loon, Lorscheid et~al.}]{berger2024towards}
\bibinfo{author}{Berger, U.}, \bibinfo{author}{Bell, A.},
  \bibinfo{author}{Barton, C.M.}, \bibinfo{author}{Chappin, E.},
  \bibinfo{author}{Dre{\ss}ler, G.}, \bibinfo{author}{Filatova, T.},
  \bibinfo{author}{Fronville, T.}, \bibinfo{author}{Lee, A.},
  \bibinfo{author}{van Loon, E.}, \bibinfo{author}{Lorscheid, I.}, et~al.,
  \bibinfo{year}{2024}.
\newblock \bibinfo{title}{Towards reusable building blocks for agent-based
  modelling and theory development}.
\newblock \bibinfo{journal}{Environmental Modelling \& Software}
  \bibinfo{volume}{175}, \bibinfo{pages}{106003}.
\newblock \DOIprefix\doi{10.1016/j.envsoft.2024.106003}.
\bibitem[{Chandrasekhar and Farimani(2025)}]{chandrasekhar2025automating}
\bibinfo{author}{Chandrasekhar, A.}, \bibinfo{author}{Farimani, A.B.},
  \bibinfo{year}{2025}.
\newblock \bibinfo{title}{Automating {MD} simulations for proteins using large
  language models: {NAMD-Agent}}.
\newblock \bibinfo{journal}{arXiv preprint} ,
  \bibinfo{pages}{2507.07887}\DOIprefix\doi{10.48550/arXiv.2507.07887},
  \href{http://arxiv.org/abs/2507.07887}{{\tt arXiv:2507.07887}}.
\bibitem[{Charness et~al.(2025)Charness, Jabarian and List}]{charness2025next}
\bibinfo{author}{Charness, G.}, \bibinfo{author}{Jabarian, B.},
  \bibinfo{author}{List, J.A.}, \bibinfo{year}{2025}.
\newblock \bibinfo{title}{The next generation of experimental research with
  {LLM}s}.
\newblock \bibinfo{journal}{Nature Human Behaviour} \bibinfo{volume}{9},
  \bibinfo{pages}{345--357}.
\newblock \DOIprefix\doi{10.1038/s41562-025-02137-1}.
\bibitem[{Chen et~al.(2021)Chen, Tworek, Jun, Yuan, de~Oliveira~Pinto, Kaplan,
  Edwards, Burda, Joseph et~al.}]{chen2021evaluating}
\bibinfo{author}{Chen, M.}, \bibinfo{author}{Tworek, J.}, \bibinfo{author}{Jun,
  H.}, \bibinfo{author}{Yuan, Q.}, \bibinfo{author}{de~Oliveira~Pinto, H.P.},
  \bibinfo{author}{Kaplan, J.}, \bibinfo{author}{Edwards, H.},
  \bibinfo{author}{Burda, Y.}, \bibinfo{author}{Joseph, N.}, et~al.,
  \bibinfo{year}{2021}.
\newblock \bibinfo{title}{Evaluating large language models trained on code}.
\newblock \bibinfo{journal}{arXiv preprint} ,
  \bibinfo{pages}{2107.03374}\DOIprefix\doi{10.48550/arXiv.2107.03374}.
\bibitem[{Cyre et~al.(1995)Cyre, Armstrong and Honcharik}]{cyre1995generating}
\bibinfo{author}{Cyre, W.R.}, \bibinfo{author}{Armstrong, J.},
  \bibinfo{author}{Honcharik, A.J.}, \bibinfo{year}{1995}.
\newblock \bibinfo{title}{Generating simulation models from natural language
  specifications}.
\newblock \bibinfo{journal}{Simulation} \bibinfo{volume}{65},
  \bibinfo{pages}{239--251}.
\newblock \DOIprefix\doi{10.1177/003754979506500402}.
\bibitem[{David et~al.(2017)David, Fachada and Rosa}]{david2017vvs}
\bibinfo{author}{David, N.}, \bibinfo{author}{Fachada, N.},
  \bibinfo{author}{Rosa, A.C.}, \bibinfo{year}{2017}.
\newblock \bibinfo{title}{Verifying and Validating Simulations}.
  \bibinfo{publisher}{Springer International Publishing},
  \bibinfo{address}{Cham}. chapter~\bibinfo{chapter}{9}.
\newblock pp. \bibinfo{pages}{173--204}.
\newblock \DOIprefix\doi{10.1007/978-3-319-66948-9\_9}.
\bibitem[{Dhruv and Dubey(2025)}]{dhruv2025leveraging}
\bibinfo{author}{Dhruv, A.}, \bibinfo{author}{Dubey, A.}, \bibinfo{year}{2025}.
\newblock \bibinfo{title}{Leveraging large language models for code translation
  and software development in scientific computing}, in:
  \bibinfo{booktitle}{Proceedings of the Platform for Advanced Scientific
  Computing Conference}, \bibinfo{publisher}{ACM}, \bibinfo{address}{New York,
  NY, USA}. pp. \bibinfo{pages}{1--9}.
\newblock \DOIprefix\doi{10.1145/3732775.3733572}.
\bibitem[{Donkin et~al.(2017)Donkin, Dennis, Ustalakov, Warren and
  Clare}]{donkin2017replicating}
\bibinfo{author}{Donkin, E.}, \bibinfo{author}{Dennis, P.},
  \bibinfo{author}{Ustalakov, A.}, \bibinfo{author}{Warren, J.},
  \bibinfo{author}{Clare, A.}, \bibinfo{year}{2017}.
\newblock \bibinfo{title}{Replicating complex agent based models, a formidable
  task}.
\newblock \bibinfo{journal}{Environmental Modelling \& Software}
  \bibinfo{volume}{92}, \bibinfo{pages}{142--151}.
\newblock \DOIprefix\doi{10.1016/j.envsoft.2017.01.020}.
\bibitem[{Edmonds et~al.(2019)Edmonds, Le~Page, Bithell, Chattoe-Brown, Grimm,
  Meyer, Monta\~{n}ola Sales, Ormerod, Root and
  Squazzoni}]{edmonds2019different}
\bibinfo{author}{Edmonds, B.}, \bibinfo{author}{Le~Page, C.},
  \bibinfo{author}{Bithell, M.}, \bibinfo{author}{Chattoe-Brown, E.},
  \bibinfo{author}{Grimm, V.}, \bibinfo{author}{Meyer, R.},
  \bibinfo{author}{Monta\~{n}ola Sales, C.}, \bibinfo{author}{Ormerod, P.},
  \bibinfo{author}{Root, H.}, \bibinfo{author}{Squazzoni, F.},
  \bibinfo{year}{2019}.
\newblock \bibinfo{title}{Different modelling purposes}.
\newblock \bibinfo{journal}{Journal of Artificial Societies and Social
  Simulation} \bibinfo{volume}{22}, \bibinfo{pages}{6}.
\newblock \DOIprefix\doi{10.18564/jasss.3993}.
\bibitem[{Elbasheer et~al.(2025)Elbasheer, Laili, Longo, Solina, Tao, Veltri,
  Zhang and Zhang}]{elbasheer2025natural}
\bibinfo{author}{Elbasheer, M.}, \bibinfo{author}{Laili, Y.},
  \bibinfo{author}{Longo, F.}, \bibinfo{author}{Solina, V.},
  \bibinfo{author}{Tao, Y.}, \bibinfo{author}{Veltri, P.},
  \bibinfo{author}{Zhang, Y.}, \bibinfo{author}{Zhang, L.},
  \bibinfo{year}{2025}.
\newblock \bibinfo{title}{Natural language-driven production planning:
  integrating large language models with automatic simulation model generation
  in manufacturing systems}.
\newblock \bibinfo{journal}{Journal of Intelligent Manufacturing}
  \DOIprefix\doi{10.1007/s10845-025-02732-z}.
\bibitem[{Fachada et~al.(2025)Fachada, Fernandes, Fernandes, Ferreira-Saraiva
  and Matos-Carvalho}]{fachada2025gpt41}
\bibinfo{author}{Fachada, N.}, \bibinfo{author}{Fernandes, D.},
  \bibinfo{author}{Fernandes, C.M.}, \bibinfo{author}{Ferreira-Saraiva, B.D.},
  \bibinfo{author}{Matos-Carvalho, J.P.}, \bibinfo{year}{2025}.
\newblock \bibinfo{title}{{GPT}-4.1 sets the standard in automated experiment
  design using novel {Python} libraries}.
\newblock \bibinfo{journal}{Future Internet} \bibinfo{volume}{17},
  \bibinfo{pages}{412}.
\newblock \DOIprefix\doi{10.3390/fi17090412}.
\bibitem[{Fachada et~al.(2026)Fachada, Fernandes, Fernandes and
  Matos-Carvalho}]{fachada2026suppl}
\bibinfo{author}{Fachada, N.}, \bibinfo{author}{Fernandes, D.},
  \bibinfo{author}{Fernandes, C.M.}, \bibinfo{author}{Matos-Carvalho, J.P.},
  \bibinfo{year}{2026}.
\newblock \bibinfo{title}{Supplementary material for ``{C}an large language
  models implement agent-based models? {A}n {ODD}-based replication study''}.
\newblock \bibinfo{howpublished}{Zenodo}.
\newblock \DOIprefix\doi{10.5281/zenodo.18521255}.
\bibitem[{Fachada et~al.(2015)Fachada, Lopes, Martins and
  Rosa}]{fachada2015template}
\bibinfo{author}{Fachada, N.}, \bibinfo{author}{Lopes, V.V.},
  \bibinfo{author}{Martins, R.C.}, \bibinfo{author}{Rosa, A.C.},
  \bibinfo{year}{2015}.
\newblock \bibinfo{title}{Towards a standard model for research in agent-based
  modeling and simulation}.
\newblock \bibinfo{journal}{PeerJ Computer Science} \bibinfo{volume}{1},
  \bibinfo{pages}{e36}.
\newblock \DOIprefix\doi{10.7717/peerj-cs.36}.
\bibitem[{Fachada et~al.(2017a)Fachada, Lopes, Martins and
  Rosa}]{fachada2017model}
\bibinfo{author}{Fachada, N.}, \bibinfo{author}{Lopes, V.V.},
  \bibinfo{author}{Martins, R.C.}, \bibinfo{author}{Rosa, A.C.},
  \bibinfo{year}{2017}a.
\newblock \bibinfo{title}{Model-independent comparison of simulation output}.
\newblock \bibinfo{journal}{Simulation Modelling Practice and Theory}
  \bibinfo{volume}{72}, \bibinfo{pages}{131--149}.
\newblock \DOIprefix\doi{10.1016/j.simpat.2016.12.013}.
\bibitem[{Fachada et~al.(2017b)Fachada, Lopes, Martins and
  Rosa}]{fachada2015parallelization}
\bibinfo{author}{Fachada, N.}, \bibinfo{author}{Lopes, V.V.},
  \bibinfo{author}{Martins, R.C.}, \bibinfo{author}{Rosa, A.C.},
  \bibinfo{year}{2017}b.
\newblock \bibinfo{title}{Parallelization strategies for spatial agent-based
  models}.
\newblock \bibinfo{journal}{International Journal of Parallel Programming}
  \bibinfo{volume}{45}, \bibinfo{pages}{449--481}.
\newblock \DOIprefix\doi{10.1007/s10766-015-0399-9}.
\bibitem[{Fachada and Rosa(2017)}]{fachada2017assessing}
\bibinfo{author}{Fachada, N.}, \bibinfo{author}{Rosa, A.C.},
  \bibinfo{year}{2017}.
\newblock \bibinfo{title}{Assessing the feasibility of {OpenCL} {CPU}
  implementations for agent-based simulations}, in:
  \bibinfo{booktitle}{Proceedings of the 5th International Workshop on OpenCL},
  \bibinfo{publisher}{ACM}, \bibinfo{address}{New York, NY, USA}. pp.
  \bibinfo{pages}{4:1--4:10}.
\newblock \DOIprefix\doi{10.1145/3078155.3078174}.
\bibitem[{Fernandes et~al.(2025)Fernandes, Matos-Carvalho, Fernandes and
  Fachada}]{fernandes2025deepseek}
\bibinfo{author}{Fernandes, D.}, \bibinfo{author}{Matos-Carvalho, J.P.},
  \bibinfo{author}{Fernandes, C.M.}, \bibinfo{author}{Fachada, N.},
  \bibinfo{year}{2025}.
\newblock \bibinfo{title}{{DeepSeek-V3}, {GPT-4}, {Phi-4}, and {LLaMA-3.3}
  generate correct code for {LoRaWAN}-related engineering tasks}.
\newblock \bibinfo{journal}{Electronics} \bibinfo{volume}{14},
  \bibinfo{pages}{1428}.
\newblock \DOIprefix\doi{10.3390/electronics14071428}.
\bibitem[{Filatova et~al.(2013)Filatova, Verburg, Parker and
  Stannard}]{filatova2013spatial}
\bibinfo{author}{Filatova, T.}, \bibinfo{author}{Verburg, P.H.},
  \bibinfo{author}{Parker, D.C.}, \bibinfo{author}{Stannard, C.A.},
  \bibinfo{year}{2013}.
\newblock \bibinfo{title}{Spatial agent-based models for socio-ecological
  systems: Challenges and prospects}.
\newblock \bibinfo{journal}{Environmental Modelling \& Software}
  \bibinfo{volume}{45}, \bibinfo{pages}{1--7}.
\newblock \DOIprefix\doi{10.1016/j.envsoft.2013.03.017}.
\bibitem[{Frydenlund et~al.(2024)Frydenlund, Mart{\'\i}nez, Padilla, Palacio
  and Shuttleworth}]{frydenlund2024modeler}
\bibinfo{author}{Frydenlund, E.}, \bibinfo{author}{Mart{\'\i}nez, J.},
  \bibinfo{author}{Padilla, J.J.}, \bibinfo{author}{Palacio, K.},
  \bibinfo{author}{Shuttleworth, D.}, \bibinfo{year}{2024}.
\newblock \bibinfo{title}{Modeler in a box: how can large language models aid
  in the simulation modeling process?}
\newblock \bibinfo{journal}{Simulation} \bibinfo{volume}{100},
  \bibinfo{pages}{727--749}.
\newblock \DOIprefix\doi{10.1177/00375497241239360}.
\bibitem[{Gao et~al.(2024)Gao, Lan, Li, Yuan, Ding, Zhou, Xu and
  Li}]{gao2024large}
\bibinfo{author}{Gao, C.}, \bibinfo{author}{Lan, X.}, \bibinfo{author}{Li, N.},
  \bibinfo{author}{Yuan, Y.}, \bibinfo{author}{Ding, J.},
  \bibinfo{author}{Zhou, Z.}, \bibinfo{author}{Xu, F.}, \bibinfo{author}{Li,
  Y.}, \bibinfo{year}{2024}.
\newblock \bibinfo{title}{Large language models empowered agent-based modeling
  and simulation: A survey and perspectives}.
\newblock \bibinfo{journal}{Humanities and Social Sciences Communications}
  \bibinfo{volume}{11}, \bibinfo{pages}{1259}.
\newblock \DOIprefix\doi{10.1057/s41599-024-03611-3}.
\bibitem[{Ge et~al.(2025)Ge, Mei, Duan, Li, Zheng, Wang, Wang, Yao, Liu, Cai
  et~al.}]{ge2025survey}
\bibinfo{author}{Ge, Y.}, \bibinfo{author}{Mei, L.}, \bibinfo{author}{Duan,
  Z.}, \bibinfo{author}{Li, T.}, \bibinfo{author}{Zheng, Y.},
  \bibinfo{author}{Wang, Y.}, \bibinfo{author}{Wang, L.}, \bibinfo{author}{Yao,
  J.}, \bibinfo{author}{Liu, T.}, \bibinfo{author}{Cai, Y.}, et~al.,
  \bibinfo{year}{2025}.
\newblock \bibinfo{title}{A survey of vibe coding with large language models}.
\newblock \bibinfo{journal}{arXiv preprint} ,
  \bibinfo{pages}{2510.12399}\DOIprefix\doi{10.48550/arXiv.2510.12399}.
\bibitem[{Grattafiori et~al.(2024)Grattafiori, Dubey, Jauhri, Pandey, Kadian,
  Al-Dahle, Letman, Mathur, Schelten, Vaughan, Yang, Fan, Goyal, Hartshorn,
  Yang et~al.}]{grattafiori2024llama}
\bibinfo{author}{Grattafiori, A.}, \bibinfo{author}{Dubey, A.},
  \bibinfo{author}{Jauhri, A.}, \bibinfo{author}{Pandey, A.},
  \bibinfo{author}{Kadian, A.}, \bibinfo{author}{Al-Dahle, A.},
  \bibinfo{author}{Letman, A.}, \bibinfo{author}{Mathur, A.},
  \bibinfo{author}{Schelten, A.}, \bibinfo{author}{Vaughan, A.},
  \bibinfo{author}{Yang, A.}, \bibinfo{author}{Fan, A.},
  \bibinfo{author}{Goyal, A.}, \bibinfo{author}{Hartshorn, A.},
  \bibinfo{author}{Yang, A.}, et~al., \bibinfo{year}{2024}.
\newblock \bibinfo{title}{The llama 3 herd of models}.
\newblock \bibinfo{journal}{arXiv preprint} ,
  \bibinfo{pages}{2407.21783}\DOIprefix\doi{10.48550/arXiv.2407.21783}.
  \bibinfo{note}{cs.AI}.
\bibitem[{Grimm et~al.(2006)Grimm, Berger, Bastiansen, Eliassen, Ginot, Giske,
  Goss-Custard, Grand, Heinz, Huse et~al.}]{grimm2006standard}
\bibinfo{author}{Grimm, V.}, \bibinfo{author}{Berger, U.},
  \bibinfo{author}{Bastiansen, F.}, \bibinfo{author}{Eliassen, S.},
  \bibinfo{author}{Ginot, V.}, \bibinfo{author}{Giske, J.},
  \bibinfo{author}{Goss-Custard, J.}, \bibinfo{author}{Grand, T.},
  \bibinfo{author}{Heinz, S.}, \bibinfo{author}{Huse, G.}, et~al.,
  \bibinfo{year}{2006}.
\newblock \bibinfo{title}{A standard protocol for describing individual-based
  and agent-based models}.
\newblock \bibinfo{journal}{Ecological Modelling} \bibinfo{volume}{198},
  \bibinfo{pages}{115--126}.
\bibitem[{Grimm et~al.(2025)Grimm, Berger, Calabrese, Cort{\'e}s-Avizanda,
  Ferrer, Franz, Groeneveld, Hartig, Jakoby, Jovani et~al.}]{grimm2025using}
\bibinfo{author}{Grimm, V.}, \bibinfo{author}{Berger, U.},
  \bibinfo{author}{Calabrese, J.M.}, \bibinfo{author}{Cort{\'e}s-Avizanda, A.},
  \bibinfo{author}{Ferrer, J.}, \bibinfo{author}{Franz, M.},
  \bibinfo{author}{Groeneveld, J.}, \bibinfo{author}{Hartig, F.},
  \bibinfo{author}{Jakoby, O.}, \bibinfo{author}{Jovani, R.}, et~al.,
  \bibinfo{year}{2025}.
\newblock \bibinfo{title}{Using the {ODD} protocol and {NetLogo} to replicate
  agent-based models}.
\newblock \bibinfo{journal}{Ecological Modelling} \bibinfo{volume}{501},
  \bibinfo{pages}{110967}.
\newblock \DOIprefix\doi{10.1016/j.ecolmodel.2024.110967}.
\bibitem[{Grimm et~al.(2010)Grimm, Berger, DeAngelis, Polhill, Giske and
  Railsback}]{grimm2010odd}
\bibinfo{author}{Grimm, V.}, \bibinfo{author}{Berger, U.},
  \bibinfo{author}{DeAngelis, D.}, \bibinfo{author}{Polhill, J.},
  \bibinfo{author}{Giske, J.}, \bibinfo{author}{Railsback, S.},
  \bibinfo{year}{2010}.
\newblock \bibinfo{title}{The {ODD} protocol: A review and first update}.
\newblock \bibinfo{journal}{Ecological Modelling} \bibinfo{volume}{221},
  \bibinfo{pages}{2760--2768}.
\newblock \DOIprefix\doi{10.1016/j.ecolmodel.2010.08.019}.
\bibitem[{Grimm et~al.(2020)Grimm, Railsback, Vincenot, Berger, Gallagher,
  DeAngelis, Edmonds, Ge, Giske, Groeneveld et~al.}]{grimm2020odd}
\bibinfo{author}{Grimm, V.}, \bibinfo{author}{Railsback, S.F.},
  \bibinfo{author}{Vincenot, C.E.}, \bibinfo{author}{Berger, U.},
  \bibinfo{author}{Gallagher, C.}, \bibinfo{author}{DeAngelis, D.L.},
  \bibinfo{author}{Edmonds, B.}, \bibinfo{author}{Ge, J.},
  \bibinfo{author}{Giske, J.}, \bibinfo{author}{Groeneveld, J.}, et~al.,
  \bibinfo{year}{2020}.
\newblock \bibinfo{title}{The {ODD} protocol for describing agent-based and
  other simulation models: A second update to improve clarity, replication, and
  structural realism}.
\newblock \bibinfo{journal}{Journal of Artificial Societies and Social
  Simulation} \bibinfo{volume}{23}, \bibinfo{pages}{7}.
\newblock \DOIprefix\doi{10.18564/jasss.4259}.
\bibitem[{Guo et~al.(2025)Guo, Yang, Zhang, Song, Zhang, Xu, Zhu, Ma, Wang, Bi
  et~al.}]{guo2025deepseekr1}
\bibinfo{author}{Guo, D.}, \bibinfo{author}{Yang, D.}, \bibinfo{author}{Zhang,
  H.}, \bibinfo{author}{Song, J.}, \bibinfo{author}{Zhang, R.},
  \bibinfo{author}{Xu, R.}, \bibinfo{author}{Zhu, Q.}, \bibinfo{author}{Ma,
  S.}, \bibinfo{author}{Wang, P.}, \bibinfo{author}{Bi, X.}, et~al.,
  \bibinfo{year}{2025}.
\newblock \bibinfo{title}{{DeepSeek-R1}: Incentivizing reasoning capability in
  llms via reinforcement learning}.
\newblock \bibinfo{journal}{arXiv preprint} ,
  \bibinfo{pages}{2501.12948}\DOIprefix\doi{10.48550/arXiv.2501.12948}.
  \bibinfo{note}{cs.CL}.
\bibitem[{Harris et~al.(2020)Harris, Millman, Van Der~Walt, Gommers, Virtanen,
  Cournapeau, Wieser, Taylor, Berg, Smith et~al.}]{harris2020array}
\bibinfo{author}{Harris, C.R.}, \bibinfo{author}{Millman, K.J.},
  \bibinfo{author}{Van Der~Walt, S.J.}, \bibinfo{author}{Gommers, R.},
  \bibinfo{author}{Virtanen, P.}, \bibinfo{author}{Cournapeau, D.},
  \bibinfo{author}{Wieser, E.}, \bibinfo{author}{Taylor, J.},
  \bibinfo{author}{Berg, S.}, \bibinfo{author}{Smith, N.J.}, et~al.,
  \bibinfo{year}{2020}.
\newblock \bibinfo{title}{Array programming with {NumPy}}.
\newblock \bibinfo{journal}{Nature} \bibinfo{volume}{585},
  \bibinfo{pages}{357--362}.
\newblock \DOIprefix\doi{10.1038/s41586-020-2649-2}.
\bibitem[{Hui et~al.(2025)Hui, Yang, Cui, Yang, Liu, Zhang, Liu, Zhang, Yu, Lu,
  Dang, Fan, Zhang, Yang, Men, Huang et~al.}]{hui2025qwen25coder}
\bibinfo{author}{Hui, B.}, \bibinfo{author}{Yang, J.}, \bibinfo{author}{Cui,
  Z.}, \bibinfo{author}{Yang, J.}, \bibinfo{author}{Liu, D.},
  \bibinfo{author}{Zhang, L.}, \bibinfo{author}{Liu, T.},
  \bibinfo{author}{Zhang, J.}, \bibinfo{author}{Yu, B.}, \bibinfo{author}{Lu,
  K.}, \bibinfo{author}{Dang, K.}, \bibinfo{author}{Fan, Y.},
  \bibinfo{author}{Zhang, Y.}, \bibinfo{author}{Yang, A.},
  \bibinfo{author}{Men, R.}, \bibinfo{author}{Huang, F.}, et~al.,
  \bibinfo{year}{2025}.
\newblock \bibinfo{title}{Qwen2.5-coder technical report}.
\newblock \bibinfo{journal}{arXiv preprint} ,
  \bibinfo{pages}{2409.12186}\DOIprefix\doi{10.48550/arXiv.2409.12186}.
  \bibinfo{note}{cs.CL}.
\bibitem[{Hurst et~al.(2024)Hurst, Lerer, Goucher, Perelman, Ramesh, Clark,
  Ostrow, Welihinda, Hayes, Radford et~al.}]{hurst2024gpt}
\bibinfo{author}{Hurst, A.}, \bibinfo{author}{Lerer, A.},
  \bibinfo{author}{Goucher, A.P.}, \bibinfo{author}{Perelman, A.},
  \bibinfo{author}{Ramesh, A.}, \bibinfo{author}{Clark, A.},
  \bibinfo{author}{Ostrow, A.}, \bibinfo{author}{Welihinda, A.},
  \bibinfo{author}{Hayes, A.}, \bibinfo{author}{Radford, A.}, et~al.,
  \bibinfo{year}{2024}.
\newblock \bibinfo{title}{{GPT-4o} system card}.
\newblock \bibinfo{journal}{arXiv preprint arXiv:2410.21276}
  \DOIprefix\doi{10.48550/arXiv.2410.21276}.
\bibitem[{Jackson et~al.(2024)Jackson, Jesus~Saenz and
  Ivanov}]{jackson2024natural}
\bibinfo{author}{Jackson, I.}, \bibinfo{author}{Jesus~Saenz, M.},
  \bibinfo{author}{Ivanov, D.}, \bibinfo{year}{2024}.
\newblock \bibinfo{title}{From natural language to simulations: applying {AI}
  to automate simulation modelling of logistics systems}.
\newblock \bibinfo{journal}{International Journal of Production Research}
  \bibinfo{volume}{62}, \bibinfo{pages}{1434--1457}.
\newblock \DOIprefix\doi{10.1080/00207543.2023.2276811}.
\bibitem[{Jiang et~al.(2026)Jiang, Wang, Shen, Kim and Kim}]{jiang2026survey}
\bibinfo{author}{Jiang, J.}, \bibinfo{author}{Wang, F.}, \bibinfo{author}{Shen,
  J.}, \bibinfo{author}{Kim, S.}, \bibinfo{author}{Kim, S.},
  \bibinfo{year}{2026}.
\newblock \bibinfo{title}{A survey on large language models for code
  generation}.
\newblock \bibinfo{journal}{ACM Transactions on Software Engineering and
  Methodology} \bibinfo{volume}{35}.
\newblock \DOIprefix\doi{10.1145/3747588}.
\bibitem[{Kamath et~al.(2025)Kamath, Ferret, Pathak, Vieillard, Merhej, Perrin,
  Matejovicova, Ram{\'e}, Rivi{\`e}re et~al.}]{kamath2025gemma3}
\bibinfo{author}{Kamath, A.}, \bibinfo{author}{Ferret, J.},
  \bibinfo{author}{Pathak, S.}, \bibinfo{author}{Vieillard, N.},
  \bibinfo{author}{Merhej, R.}, \bibinfo{author}{Perrin, S.},
  \bibinfo{author}{Matejovicova, T.}, \bibinfo{author}{Ram{\'e}, A.},
  \bibinfo{author}{Rivi{\`e}re, M.}, et~al., \bibinfo{year}{2025}.
\newblock \bibinfo{title}{Gemma 3 technical report}.
\newblock \bibinfo{journal}{arXiv preprint arXiv:2503.19786}
  \DOIprefix\doi{10.48550/arXiv.2503.19786}.
\bibitem[{Lacchia(2012)}]{radon2012}
\bibinfo{author}{Lacchia, M.}, \bibinfo{year}{2012}.
\newblock \bibinfo{title}{Radon}.
\newblock \bibinfo{howpublished}{\url{https://github.com/rubik/radon}}.
\newblock \bibinfo{note}{Last access: Aug. 24, 2025}.
\bibitem[{Lehtosalo et~al.(2014)Lehtosalo, van Rossum, Levkivskyi and
  Sullivan}]{mypy2014}
\bibinfo{author}{Lehtosalo, J.}, \bibinfo{author}{van Rossum, G.},
  \bibinfo{author}{Levkivskyi, I.}, \bibinfo{author}{Sullivan, M.J.},
  \bibinfo{year}{2014}.
\newblock \bibinfo{title}{mypy - optional static typing for {P}ython}.
\newblock \bibinfo{howpublished}{\url{https://www.mypy-lang.org/}}.
\newblock \bibinfo{note}{Last access: Aug. 24, 2025}.
\bibitem[{Liu et~al.(2024)Liu, Feng, Xue, Wang, Wu, Lu, Zhao, Deng, Zhang
  et~al.}]{liu2024deepseekv3}
\bibinfo{author}{Liu, A.}, \bibinfo{author}{Feng, B.}, \bibinfo{author}{Xue,
  B.}, \bibinfo{author}{Wang, B.}, \bibinfo{author}{Wu, B.},
  \bibinfo{author}{Lu, C.}, \bibinfo{author}{Zhao, C.}, \bibinfo{author}{Deng,
  C.}, \bibinfo{author}{Zhang, C.}, et~al., \bibinfo{year}{2024}.
\newblock \bibinfo{title}{{DeepSeek-V3} technical report}.
\newblock \bibinfo{journal}{arXiv preprint} ,
  \bibinfo{pages}{2412.19437}\DOIprefix\doi{10.48550/arXiv.2412.19437}.
  \bibinfo{note}{cs.CL}.
\bibitem[{Luo et~al.(2025)Luo, Yang, Xu, Yang and Du}]{luo2025llm4srsurvey}
\bibinfo{author}{Luo, Z.}, \bibinfo{author}{Yang, Z.}, \bibinfo{author}{Xu,
  Z.}, \bibinfo{author}{Yang, W.}, \bibinfo{author}{Du, X.},
  \bibinfo{year}{2025}.
\newblock \bibinfo{title}{Llm4sr: A survey on large language models for
  scientific research}.
\newblock \URLprefix \url{https://arxiv.org/abs/2501.04306},
  \href{http://arxiv.org/abs/2501.04306}{{\tt arXiv:2501.04306}}.
\bibitem[{Maeda and Kurata(2023)}]{maeda2023automatic}
\bibinfo{author}{Maeda, K.}, \bibinfo{author}{Kurata, H.},
  \bibinfo{year}{2023}.
\newblock \bibinfo{title}{Automatic generation of {SBML} kinetic models from
  natural language texts using {GPT}}.
\newblock \bibinfo{journal}{International Journal of Molecular Sciences}
  \bibinfo{volume}{24}, \bibinfo{pages}{7296}.
\newblock \DOIprefix\doi{10.3390/ijms24087296}.
\bibitem[{Mayer et~al.(2015)Mayer, Pappert and Rose}]{mayer2015natural}
\bibinfo{author}{Mayer, T.}, \bibinfo{author}{Pappert, F.},
  \bibinfo{author}{Rose, O.}, \bibinfo{year}{2015}.
\newblock \bibinfo{title}{A natural-language-based simulation modelling
  approach}, in: \bibinfo{editor}{Rabe, M.}, \bibinfo{editor}{Clausen, U.}
  (Eds.), \bibinfo{booktitle}{Simulation in Production and Logistics},
  \bibinfo{publisher}{Fraunhofer Verlag}. pp. \bibinfo{pages}{661--670}.
\bibitem[{McCabe(1976)}]{mccabe1976complexity}
\bibinfo{author}{McCabe, T.J.}, \bibinfo{year}{1976}.
\newblock \bibinfo{title}{A complexity measure}.
\newblock \bibinfo{journal}{IEEE Transactions on Software Engineering} ,
  \bibinfo{pages}{308--320}\DOIprefix\doi{10.1109/TSE.1976.233837}.
\bibitem[{McKinney(2011)}]{mckinney2011pandas}
\bibinfo{author}{McKinney, W.}, \bibinfo{year}{2011}.
\newblock \bibinfo{title}{pandas: a foundational python library for data
  analysis and statistics}, in: \bibinfo{booktitle}{Python for High Performance
  and Scientific Computing}.
\newblock \bibinfo{note}{18 November 2011}.
\bibitem[{{Mistral AI team}(2024a)}]{mistral2024codestral}
\bibinfo{author}{{Mistral AI team}}, \bibinfo{year}{2024}a.
\newblock \bibinfo{title}{Codestral}.
\newblock \bibinfo{howpublished}{\url{https://mistral.ai/news/codestral}}.
\newblock \URLprefix \url{https://mistral.ai/news/codestral}.
  \bibinfo{note}{accessed: 2025-07-09}.
\bibitem[{{Mistral AI team}(2024b)}]{mistral2024large}
\bibinfo{author}{{Mistral AI team}}, \bibinfo{year}{2024}b.
\newblock \bibinfo{title}{Large enough}.
\newblock
  \bibinfo{howpublished}{\url{https://mistral.ai/news/mistral-large-2407}}.
\newblock \URLprefix \url{https://mistral.ai/news/mistral-large-2407}.
  \bibinfo{note}{accessed: 2025-07-09}.
\bibitem[{M{\"u}ller et~al.(2013)M{\"u}ller, Bohn, Dre{\ss}ler, Groeneveld,
  Klassert, Martin, Schl{\"u}ter, Schulze, Weise and
  Schwarz}]{muller2013describing}
\bibinfo{author}{M{\"u}ller, B.}, \bibinfo{author}{Bohn, F.},
  \bibinfo{author}{Dre{\ss}ler, G.}, \bibinfo{author}{Groeneveld, J.},
  \bibinfo{author}{Klassert, C.}, \bibinfo{author}{Martin, R.},
  \bibinfo{author}{Schl{\"u}ter, M.}, \bibinfo{author}{Schulze, J.},
  \bibinfo{author}{Weise, H.}, \bibinfo{author}{Schwarz, N.},
  \bibinfo{year}{2013}.
\newblock \bibinfo{title}{Describing human decisions in agent-based
  models--{ODD+ D}, an extension of the {ODD} protocol}.
\newblock \bibinfo{journal}{Environmental Modelling \& Software}
  \bibinfo{volume}{48}, \bibinfo{pages}{37--48}.
\newblock \DOIprefix\doi{10.1016/j.envsoft.2013.06.003}.
\bibitem[{Oman and Hagemeister(1992)}]{oman1992metrics}
\bibinfo{author}{Oman, P.}, \bibinfo{author}{Hagemeister, J.},
  \bibinfo{year}{1992}.
\newblock \bibinfo{title}{Metrics for assessing a software system's
  maintainability}, in: \bibinfo{booktitle}{Proceedings Conference on Software
  Maintenance 1992}, \bibinfo{publisher}{IEEE}. pp. \bibinfo{pages}{337--338}.
\newblock \DOIprefix\doi{10.1109/ICSM.1992.242525}.
\bibitem[{{OpenAI}(2025)}]{openai2025gpt41}
\bibinfo{author}{{OpenAI}}, \bibinfo{year}{2025}.
\newblock \bibinfo{title}{Introducing gpt‑4.1 model family}.
\newblock \bibinfo{howpublished}{\url{https://openai.com/index/gpt-4-1/}}.
\newblock \URLprefix \url{https://openai.com/index/gpt-4-1/}.
  \bibinfo{note}{accessed: 2025-07-09}.
\bibitem[{Panda et~al.(2019)Panda, Palaniappan, Xiong, Bridgeford, Mehta, Shen
  and Vogelstein}]{panda2019hyppo}
\bibinfo{author}{Panda, S.}, \bibinfo{author}{Palaniappan, S.},
  \bibinfo{author}{Xiong, J.}, \bibinfo{author}{Bridgeford, E.W.},
  \bibinfo{author}{Mehta, R.}, \bibinfo{author}{Shen, C.},
  \bibinfo{author}{Vogelstein, J.T.}, \bibinfo{year}{2019}.
\newblock \bibinfo{title}{hyppo: A multivariate hypothesis testing python
  package}.
\newblock \bibinfo{journal}{arXiv preprint} ,
  \bibinfo{pages}{1907.02088}\DOIprefix\doi{10.48550/arXiv.1907.02088}.
\bibitem[{Pearce et~al.(2025)Pearce, Ahmad, Tan, Dolan-Gavitt and
  Karri}]{pearce2025asleep}
\bibinfo{author}{Pearce, H.}, \bibinfo{author}{Ahmad, B.},
  \bibinfo{author}{Tan, B.}, \bibinfo{author}{Dolan-Gavitt, B.},
  \bibinfo{author}{Karri, R.}, \bibinfo{year}{2025}.
\newblock \bibinfo{title}{Asleep at the keyboard? {A}ssessing the security of
  {GitHub} {Copilot}’s code contributions}.
\newblock \bibinfo{journal}{Communications of the ACM} \bibinfo{volume}{68},
  \bibinfo{pages}{96--105}.
\newblock \DOIprefix\doi{10.1145/3610721}.
\bibitem[{Pedregosa et~al.(2011)Pedregosa, Varoquaux, Gramfort, Michel,
  Thirion, Grisel, Blondel, Prettenhofer, Weiss, Dubourg, Vanderplas, Passos,
  Cournapeau, Brucher, Perrot and Duchesnay}]{pedregosa2011scikit}
\bibinfo{author}{Pedregosa, F.}, \bibinfo{author}{Varoquaux, G.},
  \bibinfo{author}{Gramfort, A.}, \bibinfo{author}{Michel, V.},
  \bibinfo{author}{Thirion, B.}, \bibinfo{author}{Grisel, O.},
  \bibinfo{author}{Blondel, M.}, \bibinfo{author}{Prettenhofer, P.},
  \bibinfo{author}{Weiss, R.}, \bibinfo{author}{Dubourg, V.},
  \bibinfo{author}{Vanderplas, J.}, \bibinfo{author}{Passos, A.},
  \bibinfo{author}{Cournapeau, D.}, \bibinfo{author}{Brucher, M.},
  \bibinfo{author}{Perrot, M.}, \bibinfo{author}{Duchesnay, E.},
  \bibinfo{year}{2011}.
\newblock \bibinfo{title}{Scikit-learn: Machine learning in {P}ython}.
\newblock \bibinfo{journal}{Journal of Machine Learning Research}
  \bibinfo{volume}{12}, \bibinfo{pages}{2825--2830}.
\bibitem[{Peng(2011)}]{peng2011reproducible}
\bibinfo{author}{Peng, R.D.}, \bibinfo{year}{2011}.
\newblock \bibinfo{title}{Reproducible research in computational science}.
\newblock \bibinfo{journal}{Science} \bibinfo{volume}{334},
  \bibinfo{pages}{1226--1227}.
\newblock \DOIprefix\doi{10.1126/science.1213847}.
\bibitem[{Rajput et~al.(2025)Rajput, Bonkoungou, Song, Kabore, Olatunji, Klein
  and Bissyande}]{rajput2025dynamic}
\bibinfo{author}{Rajput, P.}, \bibinfo{author}{Bonkoungou, A.A.},
  \bibinfo{author}{Song, Y.}, \bibinfo{author}{Kabore, A.K.},
  \bibinfo{author}{Olatunji, I.E.}, \bibinfo{author}{Klein, J.},
  \bibinfo{author}{Bissyande, T.}, \bibinfo{year}{2025}.
\newblock \bibinfo{title}{Dynamic stability of {LLM}-generated code}.
\newblock \bibinfo{journal}{arXiv preprint} ,
  \bibinfo{pages}{2511.07463}\DOIprefix\doi{10.48550/arXiv.2511.07463}.
\bibitem[{Roziere et~al.(2023)Roziere, Gehring, Gloeckle, Sootla, Gat, Tan,
  Adi, Liu, Sauvestre, Remez et~al.}]{roziere2023code}
\bibinfo{author}{Roziere, B.}, \bibinfo{author}{Gehring, J.},
  \bibinfo{author}{Gloeckle, F.}, \bibinfo{author}{Sootla, S.},
  \bibinfo{author}{Gat, I.}, \bibinfo{author}{Tan, X.E.}, \bibinfo{author}{Adi,
  Y.}, \bibinfo{author}{Liu, J.}, \bibinfo{author}{Sauvestre, R.},
  \bibinfo{author}{Remez, T.}, et~al., \bibinfo{year}{2023}.
\newblock \bibinfo{title}{Code llama: Open foundation models for code}.
\newblock \bibinfo{journal}{arXiv preprint arXiv:2308.12950}
  \DOIprefix\doi{10.48550/arXiv.2308.12950}.
\bibitem[{Sabra et~al.(2025)Sabra, Schmitt and Tyler}]{sabra2025assessing}
\bibinfo{author}{Sabra, A.}, \bibinfo{author}{Schmitt, O.},
  \bibinfo{author}{Tyler, J.}, \bibinfo{year}{2025}.
\newblock \bibinfo{title}{Assessing the quality and security of {AI}-generated
  code: A quantitative analysis}.
\newblock \bibinfo{journal}{arXiv preprint} ,
  \bibinfo{pages}{arXiv:2508.14727}\DOIprefix\doi{10.48550/arXiv.2508.14727},
  \href{http://arxiv.org/abs/2508.14727}{{\tt arXiv:2508.14727}}.
\bibitem[{Schulze et~al.(2017)Schulze, M{\"u}ller, Groeneveld and
  Grimm}]{schulze2017agent}
\bibinfo{author}{Schulze, J.}, \bibinfo{author}{M{\"u}ller, B.},
  \bibinfo{author}{Groeneveld, J.}, \bibinfo{author}{Grimm, V.},
  \bibinfo{year}{2017}.
\newblock \bibinfo{title}{Agent-based modelling of social-ecological systems:
  achievements, challenges, and a way forward}.
\newblock \bibinfo{journal}{Journal of Artificial Societies and Social
  Simulation} \bibinfo{volume}{20}.
\newblock \DOIprefix\doi{10.18564/jasss.3423}.
\bibitem[{Seabold and Perktold(2010)}]{seabold2010statsmodels}
\bibinfo{author}{Seabold, S.}, \bibinfo{author}{Perktold, J.},
  \bibinfo{year}{2010}.
\newblock \bibinfo{title}{Statsmodels: Econometric and statistical modeling
  with {Python}}.
\newblock \bibinfo{journal}{Proceedings of the 9th Python in Science
  Conference} ,
  \bibinfo{pages}{92--96}\DOIprefix\doi{10.25080/Majora-92bf1922-011}.
  \bibinfo{note}{austin, TX, USA, 28 June--3 July 2010}.
\bibitem[{Sim{\~o}es and Venson(2024)}]{simoes2024evaluating}
\bibinfo{author}{Sim{\~o}es, I.R.d.S.}, \bibinfo{author}{Venson, E.},
  \bibinfo{year}{2024}.
\newblock \bibinfo{title}{Evaluating source code quality with large language
  models: a comparative study}, in: \bibinfo{booktitle}{Proceedings of the
  XXIII Brazilian Symposium on Software Quality}, \bibinfo{publisher}{ACM},
  \bibinfo{address}{New York, NY, USA}. pp. \bibinfo{pages}{103--113}.
\newblock \DOIprefix\doi{10.1145/3701625.3701650}.
\bibitem[{Sousa et~al.(2025)Sousa, Santos, Sampaio and
  Bezerra}]{sousa2025comparing}
\bibinfo{author}{Sousa, E.}, \bibinfo{author}{Santos, E.},
  \bibinfo{author}{Sampaio, A.}, \bibinfo{author}{Bezerra, C.},
  \bibinfo{year}{2025}.
\newblock \bibinfo{title}{Comparing structural quality of code generated by
  {LLM}s: A static analysis of code smells}, in: \bibinfo{booktitle}{Anais do
  XXII Encontro Nacional de Inteligência Artificial e Computacional},
  \bibinfo{publisher}{SBC}, \bibinfo{address}{Porto Alegre, RS, Brazil}. pp.
  \bibinfo{pages}{391--402}.
\newblock \DOIprefix\doi{10.5753/eniac.2025.12470}.
\bibitem[{Sun et~al.(2016)Sun, Lorscheid, Millington, Lauf, Magliocca,
  Groeneveld, Balbi, Nolzen, M{\"u}ller, Schulze et~al.}]{sun2016simple}
\bibinfo{author}{Sun, Z.}, \bibinfo{author}{Lorscheid, I.},
  \bibinfo{author}{Millington, J.D.}, \bibinfo{author}{Lauf, S.},
  \bibinfo{author}{Magliocca, N.R.}, \bibinfo{author}{Groeneveld, J.},
  \bibinfo{author}{Balbi, S.}, \bibinfo{author}{Nolzen, H.},
  \bibinfo{author}{M{\"u}ller, B.}, \bibinfo{author}{Schulze, J.}, et~al.,
  \bibinfo{year}{2016}.
\newblock \bibinfo{title}{Simple or complicated agent-based models? a
  complicated issue}.
\newblock \bibinfo{journal}{Environmental Modelling \& Software}
  \bibinfo{volume}{86}, \bibinfo{pages}{56--67}.
\newblock \DOIprefix\doi{10.1016/j.envsoft.2016.09.006}.
\bibitem[{Sz{\'e}kely and Rizzo(2013)}]{szekely2013energy}
\bibinfo{author}{Sz{\'e}kely, G.J.}, \bibinfo{author}{Rizzo, M.L.},
  \bibinfo{year}{2013}.
\newblock \bibinfo{title}{Energy statistics: A class of statistics based on
  distances}.
\newblock \bibinfo{journal}{Journal of Statistical Planning and Inference}
  \bibinfo{volume}{143}, \bibinfo{pages}{1249--1272}.
\newblock \DOIprefix\doi{10.1016/j.jspi.2013.03.018}.
\bibitem[{Thiele and Grimm(2015)}]{thiele2015replicating}
\bibinfo{author}{Thiele, J.C.}, \bibinfo{author}{Grimm, V.},
  \bibinfo{year}{2015}.
\newblock \bibinfo{title}{Replicating and breaking models: good for you and
  good for ecology}.
\newblock \bibinfo{journal}{Oikos} \bibinfo{volume}{124},
  \bibinfo{pages}{691--696}.
\newblock \DOIprefix\doi{10.1111/oik.02170}.
\bibitem[{Walsh et~al.(2024)Walsh, Soldaini, Groeneveld, Lo, Arora, Bhagia, Gu,
  Huang, Jordan et~al.}]{olmo20242}
\bibinfo{author}{Walsh, P.}, \bibinfo{author}{Soldaini, L.},
  \bibinfo{author}{Groeneveld, D.}, \bibinfo{author}{Lo, K.},
  \bibinfo{author}{Arora, S.}, \bibinfo{author}{Bhagia, A.},
  \bibinfo{author}{Gu, Y.}, \bibinfo{author}{Huang, S.},
  \bibinfo{author}{Jordan, M.}, et~al., \bibinfo{year}{2024}.
\newblock \bibinfo{title}{2 olmo 2 furious}.
\newblock \bibinfo{journal}{arXiv preprint arXiv:2501.00656}
  \DOIprefix\doi{10.48550/arXiv.2501.00656}.
\bibitem[{Wang et~al.(2025)Wang, Ling, Wang, Yu, Wang, Li, Xiong and
  Zhang}]{wang2025maintaincoder}
\bibinfo{author}{Wang, Z.}, \bibinfo{author}{Ling, R.}, \bibinfo{author}{Wang,
  C.}, \bibinfo{author}{Yu, Y.}, \bibinfo{author}{Wang, S.},
  \bibinfo{author}{Li, Z.}, \bibinfo{author}{Xiong, F.},
  \bibinfo{author}{Zhang, W.}, \bibinfo{year}{2025}.
\newblock \bibinfo{title}{Maintaincoder: Maintainable code generation under
  dynamic requirements}.
\newblock \bibinfo{journal}{arXiv preprint} ,
  \bibinfo{pages}{2503.24260}\DOIprefix\doi{10.48550/arXiv.2503.24260}.
  \bibinfo{note}{cs.SE}.
\bibitem[{Wilensky(1997)}]{wilensky1997wolfsheep}
\bibinfo{author}{Wilensky, U.}, \bibinfo{year}{1997}.
\newblock \bibinfo{title}{{NetLogo} wolf sheep predation model}.
\newblock
  \bibinfo{howpublished}{\url{http://ccl.northwestern.edu/netlogo/models/WolfSheepPredation}}.
\bibitem[{Wilensky(1999)}]{wilensky1999netlogo}
\bibinfo{author}{Wilensky, U.}, \bibinfo{year}{1999}.
\newblock \bibinfo{title}{{NetLogo}}.
\newblock \bibinfo{howpublished}{\url{http://ccl.northwestern.edu/netlogo/}}.
\bibitem[{{xAI}(2025)}]{xai2025grok3beta}
\bibinfo{author}{{xAI}}, \bibinfo{year}{2025}.
\newblock \bibinfo{title}{Grok 3 beta --- the age of reasoning agents}.
\newblock \bibinfo{howpublished}{\url{https://x.ai/news/grok-3}}.
\newblock \URLprefix \url{https://x.ai/news/grok-3}. \bibinfo{note}{accessed:
  2025-07-09}.
\bibitem[{Yang et~al.(2025)Yang, Li, Yang, Zhang, Hui, Zheng, Yu, Gao, Huang,
  Lv et~al.}]{yang2025qwen3}
\bibinfo{author}{Yang, A.}, \bibinfo{author}{Li, A.}, \bibinfo{author}{Yang,
  B.}, \bibinfo{author}{Zhang, B.}, \bibinfo{author}{Hui, B.},
  \bibinfo{author}{Zheng, B.}, \bibinfo{author}{Yu, B.}, \bibinfo{author}{Gao,
  C.}, \bibinfo{author}{Huang, C.}, \bibinfo{author}{Lv, C.}, et~al.,
  \bibinfo{year}{2025}.
\newblock \bibinfo{title}{Qwen3 technical report}.
\newblock \bibinfo{journal}{arXiv preprint arXiv:2505.09388}
  \DOIprefix\doi{10.48550/arXiv.2505.09388}.
\bibitem[{Yeo et~al.(2024)Yeo, Ma, Kim, Jun and Kim}]{yeo2024framework}
\bibinfo{author}{Yeo, S.}, \bibinfo{author}{Ma, Y.S.}, \bibinfo{author}{Kim,
  S.C.}, \bibinfo{author}{Jun, H.}, \bibinfo{author}{Kim, T.},
  \bibinfo{year}{2024}.
\newblock \bibinfo{title}{Framework for evaluating code generation ability of
  large language models}.
\newblock \bibinfo{journal}{ETRI Journal} \bibinfo{volume}{46},
  \bibinfo{pages}{106--117}.
\newblock \DOIprefix\doi{10.4218/etrij.2023-0357}.
\bibitem[{Yildiz and Peterka(2025)}]{yildiz2025large}
\bibinfo{author}{Yildiz, O.}, \bibinfo{author}{Peterka, T.},
  \bibinfo{year}{2025}.
\newblock \bibinfo{title}{Do large language models speak scientific
  workflows?}, in: \bibinfo{booktitle}{Proceedings of the SC '25 Workshops of
  the International Conference for High Performance Computing, Networking,
  Storage and Analysis}, \bibinfo{publisher}{ACM}, \bibinfo{address}{New York,
  NY, USA}. pp. \bibinfo{pages}{2225--2233}.
\newblock \DOIprefix\doi{10.1145/3731599.3767578}.
\bibitem[{Zhao et~al.(2024)Zhao, Hui, Howland, Nguyen, Zuo, Hu, Choquette-Choo,
  Shen, Kelley et~al.}]{zhao2024codegemma}
\bibinfo{author}{Zhao, H.}, \bibinfo{author}{Hui, J.},
  \bibinfo{author}{Howland, J.}, \bibinfo{author}{Nguyen, N.},
  \bibinfo{author}{Zuo, S.}, \bibinfo{author}{Hu, A.},
  \bibinfo{author}{Choquette-Choo, C.A.}, \bibinfo{author}{Shen, J.},
  \bibinfo{author}{Kelley, J.}, et~al., \bibinfo{year}{2024}.
\newblock \bibinfo{title}{Codegemma: Open code models based on gemma}.
\newblock \bibinfo{journal}{arXiv preprint arXiv:2406.11409}
  \DOIprefix\doi{10.48550/arXiv.2406.11409}.
\bibitem[{Zhu et~al.(2024)Zhu, Guo, Shao, Yang, Wang, Xu, Wu, Li, Gao, Ma
  et~al.}]{zhu2024deepseek}
\bibinfo{author}{Zhu, Q.}, \bibinfo{author}{Guo, D.}, \bibinfo{author}{Shao,
  Z.}, \bibinfo{author}{Yang, D.}, \bibinfo{author}{Wang, P.},
  \bibinfo{author}{Xu, R.}, \bibinfo{author}{Wu, Y.}, \bibinfo{author}{Li, Y.},
  \bibinfo{author}{Gao, H.}, \bibinfo{author}{Ma, S.}, et~al.,
  \bibinfo{year}{2024}.
\newblock \bibinfo{title}{Deepseek-coder-v2: Breaking the barrier of
  closed-source models in code intelligence}.
\newblock \bibinfo{journal}{arXiv preprint arXiv:2406.11931}
  \DOIprefix\doi{10.48550/arXiv.2406.11931}.

\end{thebibliography}

\end{document}